\documentclass[final,3p,times]{elsarticle}
\usepackage[latin9]{inputenc}
\usepackage{bm}
\usepackage{amsmath}
\usepackage{amssymb}
\usepackage{graphicx}
\usepackage{esint}
\usepackage{multirow}
\usepackage[subfigure]{graphfig}
\usepackage{rotating}
\usepackage{verbatim}
\usepackage{setspace}
\usepackage[dvipsnames]{xcolor}
\makeatletter
\newcommand*{\rom}[1]{\expandafter\@slowromancap\romannumeral #1@}

\usepackage{tabularx}\usepackage{subfigure}\usepackage{subfigure}\usepackage{color}

\usepackage{setspace}

\usepackage{bm}

\newfont{\numerikEleven}{ecrm1000}
\newfont{\numerikTen}{cmss10}
\newfont{\numerikNine}{cmss9}
\newfont{\numerikEight}{cmss8}
\newfont{\numerikSeven}{cmss7}
\newfont{\numerikSix}{cmss6}

\journal{}

\makeatother

\color{black}

\begin{document}
\title{Constructing high-order discontinuity-capturing schemes with linear-weight polynomials and boundary variation diminishing algorithm}

\author[ad1]{Xi Deng \corref{cor}}

\author[ad2]{Yuya Shimizu}

\author[ad2]{Feng Xiao \corref{cor}}

\address[ad1]{Aix Marseille Univ, CNRS, Centrale Marseille, M2P2, Marseille, France}

\address[ad2]{Department of Mechanical Engineering, Tokyo Institute of Technology, 2-12-1 Ookayama, Meguro-ku, Tokyo, Japan.}

\cortext[cor]{Corresponding author: 
Dr. X. Deng (Email: deng.xi98@gmail.com), Dr. F. Xiao (Email: xiao.f.aa@m.titech.ac.jp)
}

\begin{abstract}
In this study, a new framework of constructing very high order discontinuity-capturing schemes is proposed for finite volume method. These schemes, so-called $\mathrm{P}_{n}\mathrm{T}_{m}-\mathrm{BVD}$  (polynomial of $n$-degree and THINC function of $m$-level reconstruction based on BVD algorithm), are designed by employing high-order linear-weight polynomials and THINC (Tangent of Hyperbola for INterface Capturing) functions with adaptive steepness as the reconstruction candidates. The final reconstruction function in each cell is determined with a multi-stage BVD (Boundary Variation Diminishing) algorithm so as to effectively control numerical oscillation and dissipation. We devise the new schemes up to eleventh order in an efficient way by directly increasing the order of the underlying upwind scheme using linear-weight polynomials. The analysis of the spectral property and accuracy tests show that the new reconstruction strategy well preserves the low-dissipation property of the underlying upwind schemes with high-order linear-weight polynomials for smooth solution over all wave numbers and realizes $n+1$ order convergence rate. The performance of new schemes is examined through widely used benchmark tests, which demonstrate that the proposed schemes are capable of  simultaneously resolving small-scale flow features with high resolution and capturing discontinuities with low dissipation. With outperforming results and simplicity in algorithm, the new reconstruction strategy shows great potential as an alternative numerical framework for computing nonlinear hyperbolic conservation laws that have discontinuous and smooth solutions of different scales.

\end{abstract}

\maketitle

\section{Introduction}
Designing shock-capturing schemes for high speed compressible flows involving complex flow structures of wide range scales still remains an unresolved issue and attracts a lot of attention in the computational fluid dynamics community. The coexistence of discontinuities and small-scale flow features poses big challenge to existing numerical schemes. High order and low dissipative schemes are demanded to resolve high-frequency waves and vortices featured in turbulent flows. However, high order schemes may cause spurious numerical oscillations when solving discontinuities such as shock waves, contacts and material interfaces. With the order increased, a scheme may become less robust or too oscillatory to use. To suppress numerical oscillations associated with high order interpolation, a common practice is to introduce certain amount of numerical dissipations by projecting high order interpolation polynomials to lower order or smoother ones, which is also known as limiting projection. An ideal limiting projection is expected to have as small as possible numerical dissipation, while still to effectively suppress numerical oscillation around critical regions. Excessive numerical dissipation introduced in limiting processes tends to undermine the accuracy even through high order interpolation is employed.    

Over the decades, a great deal of efforts has been made to construct high order shock capturing schemes. TVD (Total Variation Diminishing) schemes \cite{harten-tvd}, such as the MUSCL (Monotone Upstream-centered Schemes for Conservation Law) scheme \cite{Van_Leer}, can resolve discontinuities without numerical oscillations by introducing slope or flux limiters, which are usually formulated in a solution-dependent or nonlinear fashion to degrade the reconstruction function down to a piecewise constant interpolation to ensure the TVD property in vicinity of discontinuities. However, although TVD schemes can ensure the physical fields to be bounded and monotonic in the transition region, they typically suffer from excessive numerical dissipation. To reduce numerical dissipation, ENO (Essentially Non-oscillatory) schemes have been proposed to use higher-order polynomials with a less restrictive requirement in the monotonicity of numerical solutions \cite{Harten1,Harten2,shu_eno1,shu_eno2}. The basic idea of ENO is to construct several polynomial interpolations with different candidate stencils and to choose the smoothest one as the final reconstruction function. 
Following the ENO schemes, WENO (Weighted Essentially Non-oscillatory) schemes have been devised in \cite{liu94,jiang96}. Instead of choosing the smoothest stencil, WENO schemes are built on a weighted average of approximations from all candidate stencils. The nonlinar weight of each stencil is assigned according to smoothness indicators which depends on the numerical solution. In the vicinity of discontinuities, the ENO property is realized through assigning small weights to less smooth candidates. In smooth regions, weights are designed so as to restore the reconstruction function to the highest possible underlying polynomial of linear weights\footnote{We refer to in the context the schemes that use polynomials of linear weights as linear schemes, and those using nonlinear weights as nonlinear schemes. }.  Thus, using whole candidate stencils, WENO schemes are more accurate in smooth region than ENO schemes.

The WENO scheme proposed in \cite{jiang96}, known as the classical WENO scheme, achieves high order accuracy for smooth regions and are essentially non-oscillatory near discontinuities. Despite of its success, it has been recognized in \cite{WENOM} that the classical WENO scheme generates excessive numerical dissipation that tends to smear out contact discontinuities or jumps in variables across material interface. Moreover, the numerical dissipation introduced by the classical WENO scheme significantly damps the small-scale flow structures for turbulence simulations \cite{TENO14,TENO15}. Since then numerous studies have been contributed to further improve the accuracy of WENO schemes. For example, a series of new smoothness indicators have been proposed in \cite{wenoz,wenoh,wenop,wenozn} where contributions of the less smooth candidate stencils are optimized to reduce numerical dissipation. To improve the accuracy for turbulence flow, in \cite{wenohu} numerical dissipation is reduced through employing central discretization. It is observed that the spectral property of the WENO schemes is inferior to that of low-dissipation linear scheme with high-order polynomials. Some recent efforts have been made to improve the spectral property of the WENO schemes. In \cite{TENO}, a family of high order targeted ENO (TENO) scheme has been proposed which is capable of preserving low dissipation of linear schemes for low and intermediate wavenumber region. In \cite{embedded}, a new design strategy named embedded WENO is devised to utilise all adjacent smooth substencils to construct a desirable interpolation thus to improve the behavior of the spectral property. In spite of these works, there is almost no nonlinear WENO adaptation in the existing schemes which can completely retrieve the spectral property of the underlying linear schemes in high wavenumber regime. Thus WENO nonlinear adaptation introduces excessive numerical dissipation to smear out small-scale flow structures. The multi-dimensional optimal order detection (MOOD) method \cite{raphael11,raphael12,raphael13} provides a framework to hybrid higher-order linear schemes with shock capturing schemes, which enables to realize the spectral property of the linear schemes through posteriori corrections and to show superiority in resolving multi-scale structures. 

Besides improving the smoothness indicator of WENO, another approach to reduce numerical dissipation is to increase the order of reconstruction function. Very high order WENO schemes have been also devised and studied systematically in \cite{highWENO,veryhigh}, which shows that increasing order of WENO scheme effectively reduces the numerical dissipation and improves numerical results in turbulence flow simulations. However, as reported in these works, oscillations will grow and contaminate the solutions with increased order for reconstruction since higher-order reconstruction may be unlikely to find a wide enough smooth stencil. Thus in \cite{veryhigh}, a strategy that recursively reduces the order when a reconstruction-failure is detected is adopted to prevent oscillations. This additional strategy further increases the complexity of designing very high order scheme. Legendre polynomials were suggested in \cite{WENOAO} as efficient and economical option to  construct very high order WENO schemes.  

We have recently developed a new class of shock-capturing schemes \cite{Sun,dengCF,deng2018a,deng2018b,deng2019} which hybrid polynomial based nonlinear reconstructions and the THINC (Tangent of Hyperbola for INterface Capturing) function as the reconstruction candidates. Realizing that polynomial-based reconstruction may not be suitable for discontinuities, in \cite{Sun} a non-polynomial jump-like THINC function is employed to represent the discontinuity. The switch between WENO and THINC is guided by the BVD (Boundary Variation Diminishing) algorithms, which select reconstruction functions by minimizing the jumps of the reconstructed values at the cell boundaries. The spectral analysis and numerical experiments \cite{dengCF} reveal that THINC function with properly chosen steepness can outperform most existing MUSCL schemes for spatial reconstruction. A new scheme, the adaptive THINC-BVD scheme,  was then designed under the BVD principle by employing THINC functions with different sharpnesses to solve both smooth and discontinuous solutions \cite{dengCF}. Since THINC is a monotonic and bounded function, the adaptive THINC-BVD scheme is able to eliminate numerical oscillation without any limiting projection. For better use of the numerical property of the linear-weight polynomial function, a fifth-order scheme is devised by employing a linear fifth-order upwind scheme and THINC function with the BVD algorithm. The resulting scheme can retrieve linear upwind scheme in smooth region and capture sharp discontinuities without numerical oscillations. Our practice so far shows that the BVD principle can be used as a general and effective paradigm to design new numerical schemes.

In this paper, we propose a new general framework to construct high order shock capture scheme. These schemes, so-called $\mathrm{P}_{n}\mathrm{T}_{m}-\mathrm{BVD}$ (polynomial of $n$-degree and THINC function of $m$-level reconstruction based on BVD algorithm), are designed by employing $n+1$ order upwind schemes with linear-weight polynomials and THINC (Tangent of Hyperbola for INterface Capturing) functions with adaptive steepness as reconstruction candidates. The final reconstruction function in each cell is determined with multi-stage BVD (Boundary Variation Diminishing) algorithm so as to effectively control numerical oscillation and dissipation. As shown later in the paper, the proposed $\mathrm{P}_{n}\mathrm{T}_{m}-\mathrm{BVD}$ schemes have following desirable properties: 1) they solve sharp discontinuities with effectively suppressed numerical oscillations; 2) they retrieve the underlying linear high-order schemes for smooth solution over all wave numbers, and thus substantially reduce numerical dissipation errors for small-scale flow structures; 3) they can be extended to high order in an efficient way by simply increasing the order of the underlying upwind scheme.

The remainder of this paper is organized as follows. In Section 2, after a brief review of the finite volume method, the details of the new scheme for spatial reconstruction are presented. The spectral property of the new scheme will also be presented. In Section 3, the performance of the new scheme will be examined through benchmark tests in 1D and 2D. Some concluding remarks and prospective are given in Section 4.

\section{Numerical methods \label{sec:model}}

\subsection{Finite volume method}
In this paper, the 1D scalar conservation law in following form is used to introduce the new scheme
\begin{equation}
\label{eq:scalar}
\frac{\partial q}{\partial t} + \frac{\partial f(q)}{\partial x} = 0,
\end{equation}
where $q(x,t)$ is the solution function and $f(q)$ is the flux function. We divide the computational domain into $N$ non-overlapping cell elements, ${\mathcal I}_{i}: x \in [x_{i-1/2},x_{i+1/2} ]$, $i=1,2,\ldots,N$, with a uniform grid spacing $h=\Delta x=x_{i+1/2}-x_{i-1/2}$. For a standard finite volume method, the volume-integrated average value $\bar{q}_{i}(t)$ in cell ${\mathcal I}_{i}$ is defined as
\begin{equation}
\bar{q}_{i}(t) 
\approx \frac{1}{\Delta x} \int_{x_{i-1/2}}^{x_{i+1/2}}
q(x,t) \; dx.
\end{equation}
The semi-discrete version of Eq.~(\ref{eq:scalar}) in the finite volume form can be expressed as an ordinary differential equation (ODE)
\begin{equation}
\frac{\partial \bar{q}(t)}{\partial t}  =-\frac{1}{\Delta x}(\tilde{f}_{i+1/2}-\tilde{f}_{i-1/2}),
\end{equation}
where the numerical fluxes $\tilde{f}$ at cell boundaries can be computed by a Riemann solver
\begin{equation}
\tilde{f}_{i+1/2}=f_{i+1/2}^{\text{Riemann}}(q_{i+1/2}^{L},q_{i+1/2}^{R})
\end{equation}
as long as the reconstructed left-side value $q_{i+1/2}^{L}$ and right-side value $q_{i+1/2}^{R}$ at cell boundaries are provided. Essentially, the Riemann flux can be written in a canonical form as
\begin{equation}
\label{eq:Riemann}
f_{i+1/2}^{\text{Riemann}}(q_{i+1/2}^{L},q_{i+1/2}^{R}) =\frac{1}{2}\left(f(q_{i+1/2}^{L})+f(q_{i+1/2}^{R})\right)-\frac{|a_{i+1/2}|}{2}\left(q_{i+1/2}^{R}-q_{i+1/2}^{L})\right),
\end{equation}
where $a_{i+1/2}$ stands for the characteristic speed of the hyperbolic conservation law. The remaining main task is how to calculate $q_{i+1/2}^{L}$ and $q_{i+1/2}^{R}$ through the reconstruction process.

\subsection{Reconstruction process}
In this subsection, we give the details of how to calculate $q_{i+1/2}^{L}$ and $q_{i+1/2}^{R}$ using the BVD principle.
The proposed $\mathrm{P}_{n}\mathrm{T}_{m}-\mathrm{BVD}$ schemes are designed by employing linear weight polynomial of $n$-degree and THINC function of $m$-level as the candidate interpolants. The final reconstruction function in each cell is selected from these candidate interpolants with the BVD algorithm. Next, we introduce the candidate interpolants before the description of the BVD algorithm.

\subsubsection{Candidate interpolant $\mathrm{P}_{n}$: linear upwind scheme of $n$-degree polynomial}
A finite volume scheme of $(n+1)$th order can be constructed from a spatial approximation for the solution in the target cell ${\mathcal I}_{i}$ with a polynomial $\tilde{q}_{i}^{Pn}(x)$ of degree $n$. The $n+1$ unknown coefficients of the polynomial are determined by requiring that $\tilde{q}_{i}^{Pn}(x)$ has the same cell average on each cell over an appropriately selected stencil $S=\{i-n^{-},\dots, i+n^{+}\}$ with $n^{-}+n^{+}=n$, which is expressed as
\begin{equation}\label{Eq:linearR}
\dfrac{1}{\Delta x}\int_{x_{j-1/2}}^{x_{j+1/2}}\tilde{q}_{i}^{Pn}(x)dx= \bar{q}_{j}, ~~j=i-n^{-},i-n^{-}+1,\dots,i+n^{+}.
\end{equation}
To construct $2r-1$ order upwind-biased (UW) finite volume schemes as detailed in \cite{liu94,jiang96,very3}, the stencil is defined with $n^{-}=n^{+}=r-1$. The unknown coefficients of polynomial of $2r-2$ degree can be then calculated from \eqref{Eq:linearR}. With the polynomial $\tilde{q}_{i}^{Pn}(x)$, high order approximation for reconstructed values at the cell boundaries can be obtained by
\begin{equation}\label{Eq:upwind}
q^{L,Pn}_{i+\frac{1}{2}}=\tilde{q}_{i}^{Pn}(x_{i+\frac{1}{2}}) \ \ {\rm and} \ \  q^{R,Pn}_{i-\frac{1}{2}}=\tilde{q}_{i}^{Pn}(x_{i-\frac{1}{2}}).
\end{equation}
The analysis of \cite{liu94,jiang96,very3} shows that in smooth region, the approximation with polynomial \eqref{Eq:upwind} can achieve $2r-1$ order accuracy. 
In this work, we extend the proposed scheme from fifth order ($r=3$) to eleventh order ($r=6$) by using polynomials of 4th, 6th, 8th and 10th degree as underlying scheme for smooth solution. In order to facilitate the implementation of the proposed method, we give the explicit formulas of $q_{i+1/2}^{L,Pn}$ and $q_{i-1/2}^{R,Pn}$ for (n+1)th-order scheme as follows, 
\begin{itemize}
 \item 5th-order scheme
\begin{equation}\label{Eq:5thuw}
\begin{aligned}
q_{i+1/2}^{L,P4}=\dfrac{1}{30}\bar{q}_{i-2}-\dfrac{13}{60}\bar{q}_{i-1}+\dfrac{47}{60}\bar{q}_{i}+\dfrac{9}{20}\bar{q}_{i+1}-\dfrac{1}{20}\bar{q}_{i+2}, \\
q_{i-1/2}^{R,P4}=\dfrac{1}{30}\bar{q}_{i+2}-\dfrac{13}{60}\bar{q}_{i+1}+\dfrac{47}{60}\bar{q}_{i}+\dfrac{9}{20}\bar{q}_{i-1}-\dfrac{1}{20}\bar{q}_{i-2}.
\end{aligned}
\end{equation}
 \item 7th-order scheme 
\begin{equation}\label{Eq:7thuw}
\begin{aligned}
q_{i+1/2}^{L,P6}=-\dfrac{1}{140}\bar{q}_{i-3}+\dfrac{5}{84}\bar{q}_{i-2}-\dfrac{101}{420}\bar{q}_{i-1}+\dfrac{319}{420}\bar{q}_{i}+\dfrac{107}{210}\bar{q}_{i+1}-\dfrac{19}{210}\bar{q}_{i+2}+\dfrac{1}{105}\bar{q}_{i+3}, \\
q_{i-1/2}^{R,P6}=-\dfrac{1}{140}\bar{q}_{i+3}+\dfrac{5}{84}\bar{q}_{i+2}-\dfrac{101}{420}\bar{q}_{i+1}+\dfrac{319}{420}\bar{q}_{i}+\dfrac{107}{210}\bar{q}_{i-1}-\dfrac{19}{210}\bar{q}_{i-2}+\dfrac{1}{105}\bar{q}_{i-3}.
\end{aligned}
\end{equation}
 \item 9th-order scheme
\begin{equation}\label{Eq:9thuw}
\begin{aligned}
q_{i+1/2}^{L,P8}=\dfrac{1}{630}\bar{q}_{i-4}-\dfrac{41}{2520}\bar{q}_{i-3}+\dfrac{199}{2520}\bar{q}_{i-2}-\dfrac{641}{2520}\bar{q}_{i-1}+\dfrac{1879}{2520}\bar{q}_{i}+\dfrac{275}{504}\bar{q}_{i+1}-\dfrac{61}{504}\bar{q}_{i+2}+\dfrac{11}{501}\bar{q}_{i+3}-\dfrac{1}{504}\bar{q}_{i+4}, \\
q_{i-1/2}^{R,P8}=\dfrac{1}{630}\bar{q}_{i+4}-\dfrac{41}{2520}\bar{q}_{i+3}+\dfrac{199}{2520}\bar{q}_{i+2}-\dfrac{641}{2520}\bar{q}_{i+1}+\dfrac{1879}{2520}\bar{q}_{i}+\dfrac{275}{504}\bar{q}_{i-1}-\dfrac{61}{504}\bar{q}_{i-2}+\dfrac{11}{501}\bar{q}_{i-3}-\dfrac{1}{504}\bar{q}_{i-4}.
\end{aligned}
\end{equation}
 \item 11th-order scheme
\begin{equation}\label{Eq:11thuw}
\begin{array}{llll}
q_{i+1/2}^{L,P10}=&-\dfrac{1}{2772}\bar{q}_{i-5}+\dfrac{61}{13860}\bar{q}_{i-4}-\dfrac{703}{27720}\bar{q}_{i-3}+\dfrac{371}{3960}\bar{q}_{i-2}-\dfrac{7303}{27720}\bar{q}_{i-1}+\dfrac{20417}{27720}\bar{q}_{i}+\dfrac{15797}{27720}\bar{q}_{i+1}-\dfrac{4003}{27720}\bar{q}_{i+2}\\&-\dfrac{947}{27720}\bar{q}_{i+3}+\dfrac{17}{3080}\bar{q}_{i+4}+\dfrac{1}{2310}\bar{q}_{i+5}, \\
q_{i-1/2}^{R,P10}=&-\dfrac{1}{2772}\bar{q}_{i+5}+\dfrac{61}{13860}\bar{q}_{i+4}-\dfrac{703}{27720}\bar{q}_{i+3}+\dfrac{371}{3960}\bar{q}_{i+2}-\dfrac{7303}{27720}\bar{q}_{i+1}+\dfrac{20417}{27720}\bar{q}_{i}+\dfrac{15797}{27720}\bar{q}_{i-1}-\dfrac{4003}{27720}\bar{q}_{i-2}\\&-\dfrac{947}{27720}\bar{q}_{i-3}+\dfrac{17}{3080}\bar{q}_{i-4}+\dfrac{1}{2310}\bar{q}_{i-5}.
\end{array}
\end{equation} 
\end{itemize}

\subsubsection{Candidate interpolant $\mathrm{T}_{m}$: non-polynomial THINC function with $m$-level steepness}
Another candidate interpolation function in our scheme makes use of the THINC interpolation which is a differentiable and monotone Sigmoid function \cite{xiao_thinc,xiao_thinc2}.  
The piecewise THINC reconstruction function is written as
\begin{equation} \label{eq:THINC}
\tilde{q}_{i}^{T}(x)=\bar{q}_{min}+\dfrac{\bar{q}_{max}}{2} \left[1+\theta~\tanh \left(\beta \left(\dfrac{x-x_{i-1/2}}{x_{i+1/2}-x_{i-1/2}}-\tilde{x}_{i}\right)\right)\right],
\end{equation} 
where $\bar{q}_{min}=\min(\bar{q}_{i-1},\bar{q}_{i+1})$, $\bar{q}_{max}=\max(\bar{q}_{i-1},\bar{q}_{i+1})-\bar{q}_{min}$ and 
$\theta=sgn(\bar{q}_{i+1}-\bar{q}_{i-1})$. The jump thickness is controlled by the parameter $\beta$, i.e. a small value of $\beta$ leads to a smooth profile while a large one leads to a sharp jump-like distribution. 
The unknown $\tilde{x}_{i}$, which represents the location of the jump center, is computed from constraint condition $\displaystyle \bar{q}_{i} = \frac{1}{\Delta x} \int_{x_{i-1/2}}^{x_{i+1/2}} \tilde{q}^{T}_{i}(x) dx$. 

Since the value given by hyperbolic tangent function $\tanh(x)$ lays in the region of $[-1,1]$, 
the value of THINC reconstruction function $\tilde{q}_{i}^{T}(x)$ is rigorously bounded by $\bar{q}_{i-1}$ and $\bar{q}_{i+1}$. 
Given the reconstruction function $\tilde{q}_{i}^{T}(x)$, we calculate the boundary values $q_{i+1/2}^{L,T}$ and $q_{i-1/2}^{R,T}$ by $q_{i+1/2}^{L,T}=\tilde{q}_{i}^{T}(x_{i+1/2})$ and $q_{i-1/2}^{R,T}=\tilde{q}_{i}^{T}(x_{i-1/2})$ respectively. 

In \cite{dengCF}, the effect of the sharpness parameter $\beta$ on numerical dissipation of the THINC scheme has been investigated with approximate dispersion relation (ADR) analysis. It is concluded that (i) with $\beta=1.1$, THINC has much smaller numerical dissipation than TVD scheme with Minmod limiter \cite{books}, and has similar but slightly better performance than the Van Leer limiter\cite{Van_Leer}; (ii) with $\beta=1.2$, the dissipation property of THINC is between Van Leer and Superbee limiter; (iii) with a larger $\beta$, compressive or anti-diffusion effect will be introduced, which is preferred for discontinuous solutions. Based on this observation, a reconstruction strategy for both smooth and discontinuous solutions is  proposed by adaptively choosing the sharpness parameter $\beta$ with the BVD principle\cite{dengCF}. 

Following \cite{dengCF}, in this work we use THINC functions with $\beta$ of $m$-level to represent different steepness, to realize non-oscillatory and less-dissipative reconstructions adaptively for various flow structures.  A THINC reconstruction function $\tilde{q}_{i}^{Tk}(x)$ with $\beta_{k}$ gives the reconstructed values $q_{i+1/2}^{L,Tk}$ and $q_{i-1/2}^{R,Tk}$, ($k=1,2,\dots,m$). We will use $m$ up to three in present study.

\subsubsection{The BVD algorithm} 
The underlying high order upwind schemes \eqref{Eq:5thuw}-\eqref{Eq:11thuw} can achieve the targeted optimal order for smooth region. However, numerical oscillations will appear for discontinuous solutions, such as the shock waves in high speed compressible flow. As aforementioned, in the conventional high resolution schemes,  nonlinear limiting projections are designed to suppress such numerical oscillations in presence of discontinuous solutions. Unfortunately, these limiting processes usually undermine the accuracy and can hardly retrieve the original high-order linear schemes for smooth solutions of relatively small scales. An ideal limiting projection should maintain as much as possible the numerical properties of the scheme that uses the original polynomial of linear weights. In this work, we present a novel reconstruction scheme based on the BVD principle, where the linear high order schemes are directly used for smooth solutions. 

In \cite{dengCF}, a variant BVD algorithm was devised to minimize the total boundary variation (TBV), which implies that the reconstruction function which fits better with the flow field distribution will give a smaller boundary variation value. Thus using monotonic interpolations results in smaller boundary variation values in presence of discontinuous solution, whereas high-order interpolations are preferred to minimize the boundary variations for smooth region. In the $\mathrm{P}_{n}\mathrm{T}_{m}-\mathrm{BVD}$ schemes presented in this paper, reconstruction function is determined from the candidate interpolants with $k$-stage $(k=m)$ BVD algorithm so as to minimize the TBV of the target cell. We denote the reconstruction function in the target cell ${\mathcal I}_{i}$ after the $k$-th stage BVD as $\tilde{q}_{i}^{<k>}(x)$.  

The $k$-stage BVD algorithm is formulated as follows. 

\begin{description} 
	\item{\bf (I):  Initial stage $(k=0)$:}  
	\begin{description} 
		\item (I-I) As the first step, use the linear high-order upwind scheme as the base reconstruction scheme and initialize the reconstructed function as  $\tilde{q}_{i}^{<0>}(x)=\tilde{q}_{i}^{Pn(x)}$.  
	\end{description} 
\end{description} 

\begin{description} 
	\item{\bf (II) The intermediate BVD stage $(k=1,\dots,m-1)$: }  
	\begin{description} 
		\item (II-I) Set $\tilde{q}_{i}^{<k>}(x)=\tilde{q}_{i}^{<k-1>}(x)$
		\item (II-II) Calculate the TBV values for target cell ${\mathcal I}_{i}$ from the reconstruction of $\tilde{q}_{i}^{<k>}(x)$
		\begin{equation}\label{Eq:TBVp4}
		TBV_{i}^{<k>}=\big|q_{i-1/2}^{L,<k>}-q_{i-1/2}^{R,<k>}\big|+\big|q_{i+1/2}^{L,<k>}-q_{i+1/2}^{R,<k>} \big|
		\end{equation} 
		and from the THINC function $\tilde{q}_{i}^{Tk}(x)$ with a  steepness $\beta_{k}$ as    
		\begin{equation}\label{Eq:TBVTs}
		TBV_{i}^{Tk}=\big|q_{i-1/2}^{L,Tk}-q_{i-1/2}^{R,Tk}\big|+\big|q_{i+1/2}^{L,Tk}-q_{i+1/2}^{R,Tk} \big|.
		\end{equation} 
		\item (II-III) Modify the reconstruction function for cells $i-1$, $i$ and $i+1$ according to the following BVD algorithm
		\begin{equation}\label{Eq:BVDlim-1}
		\tilde{q}_{j}^{<k>}(x)=\tilde{q}_{j}^{Tk}(x), \ j=i-1,i,i+1;~~~{\rm if } \ \ TBV_{i}^{Tk}  < TBV_{i}^{<k>}.
		\end{equation} 
	\end{description} 
\end{description} 

\begin{description} 
	\item {\bf  (III) The final BVD stage $(k=m)$:} 
	\begin{description} 
		\item (III-I) 
		Given the reconstruction functions $\tilde{q}_{i}^{<m-1>}(x)$ from above stage, compute the TBV using the reconstructed cell boundary values from previous stage by     
		\begin{equation}\label{Eq:TBVlim}
		TBV_{i}^{<m-1>}=\big|q_{i-1/2}^{L,<m-1>}-q_{i-1/2}^{R,<m-1>}\big|+\big|q_{i+1/2}^{L,<m-1>}-q_{i+1/2}^{R,<m-1>} \big|, 
		\end{equation}
		and the TBV for THINC function of $\beta_{m}$ by 
		\begin{equation}\label{Eq:TBVtl}
		TBV_{i}^{Tm}=\big|q_{i-1/2}^{L,Tm}-q_{i-1/2}^{R,Tm}\big|+\big|q_{i+1/2}^{L,Tm}-q_{i+1/2}^{R,Tm} \big|. 
		\end{equation}
		\item (III-II)  Determine the final reconstruction function for cell ${\mathcal I}_{i}$ using the  BVD algorithm as
		\begin{equation}
		\tilde{q}_{i}^{<m>}(x)=\left\{
		\begin{array}{l}
		\tilde{q}_{i}^{Tm};~~~{\rm if } \ \ TBV_{i}^{Tm}  < TBV_{i}^{<m-1>}, \\
		\tilde{q}_{i}^{<m-1>};~~~~\mathrm{otherwise}
		\end{array}
		\right..
		\end{equation}
		
		\item (III-III) Compute the reconstructed values on the left-side of $x_{i+\frac{1}{2}}$ and the right-side of $x_{i-\frac{1}{2}}$  respectively by 
		\begin{equation}
		q^{L}_{i+\frac{1}{2}}=\tilde{q}_{i}^{<m>}(x_{i+\frac{1}{2}}) \ \ {\rm and} \ \  q^{R}_{i-\frac{1}{2}}=\tilde{q}_{i}^{<m>}(x_{i-\frac{1}{2}}).
		\end{equation}

	\end{description} 
\end{description} 

In this study, we propose and test fifth order scheme with $\mathrm{P}_{4}\mathrm{T}_{2}-\mathrm{BVD}$, seventh order scheme with $\mathrm{P}_{6}\mathrm{T}_{3}-\mathrm{BVD}$, ninth order scheme with $\mathrm{P}_{8}\mathrm{T}_{3}-\mathrm{BVD}$ and eleventh order scheme with $\mathrm{P}_{10}\mathrm{T}_{3}-\mathrm{BVD}$. According to previous study in \cite{dengCF}, in all tests of the present study we use $\beta_{1}=1.1$ and $\beta_{2}=1.8$ for $\mathrm{P}_{4}\mathrm{T}_{2}-\mathrm{BVD}$, and  $\beta_{1}=1.2$, $\beta_{2}=1.1$ and $\beta_{3}=1.8$ for $\mathrm{P}_{n}\mathrm{T}_{3}-\mathrm{BVD} (n=6,8,10)$ schemes. 

\begin{description}
	\item {Remark 1. }The multi-stage BVD algorithm enables to reinforce the desired numerical properties at different stages. In the intermediate stage $(k=1,2,\dots,m-1)$, the oscillation-free property is realized. The final stage ($k=m$) is devised to reduce numerical dissipation to capture sharp discontinuities. In the final stage, the parameter $\beta$ can be chosen from 1.6 to 2.2. A larger value will result in a sharper discontinuity. 
	\item {Remark 2. }This work provides a new framework to construct high order shock capturing schemes. As shown above, extending the scheme from 7th to 11th order is straightforward by simply applying the linear-weight polynomials of the targeted order as given in \eqref{Eq:7thuw}-\eqref{Eq:11thuw}. Our numerical experiments shows that higher-order schemes beyond 11th order can be also designed by adding more levels of THINC function and BVD algorithm in the same spirit.  
        	
\end{description}
 
\subsubsection{Spectral property}
\begin{figure}
	\begin{center}
		\subfigure{\centering\includegraphics[scale=0.35,trim={0.5cm 0.5cm 0.5cm 0.5cm},clip]{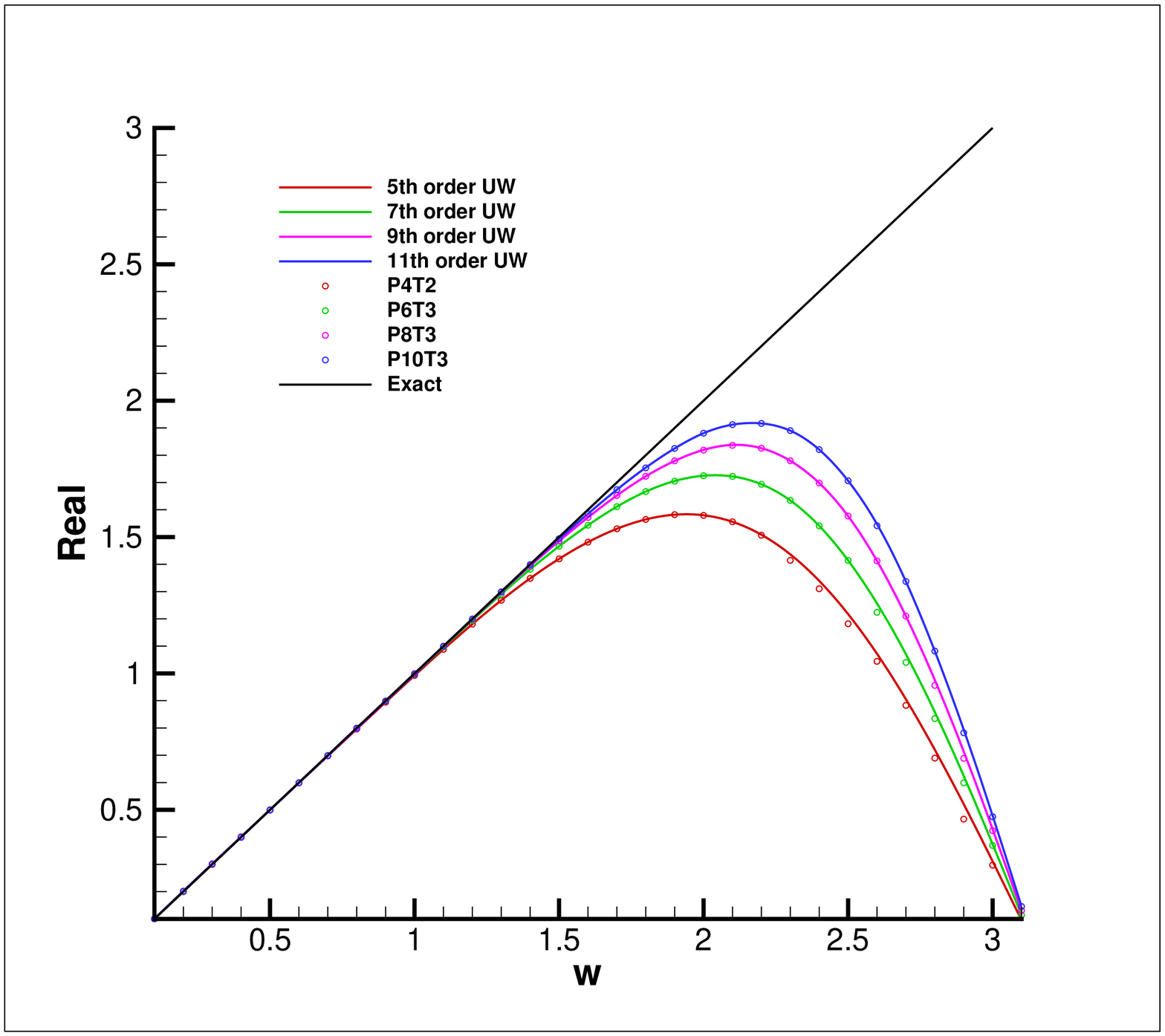}}
		\subfigure{\centering\includegraphics[scale=0.35,trim={0.5cm 0.5cm 0.5cm 0.5cm},clip]{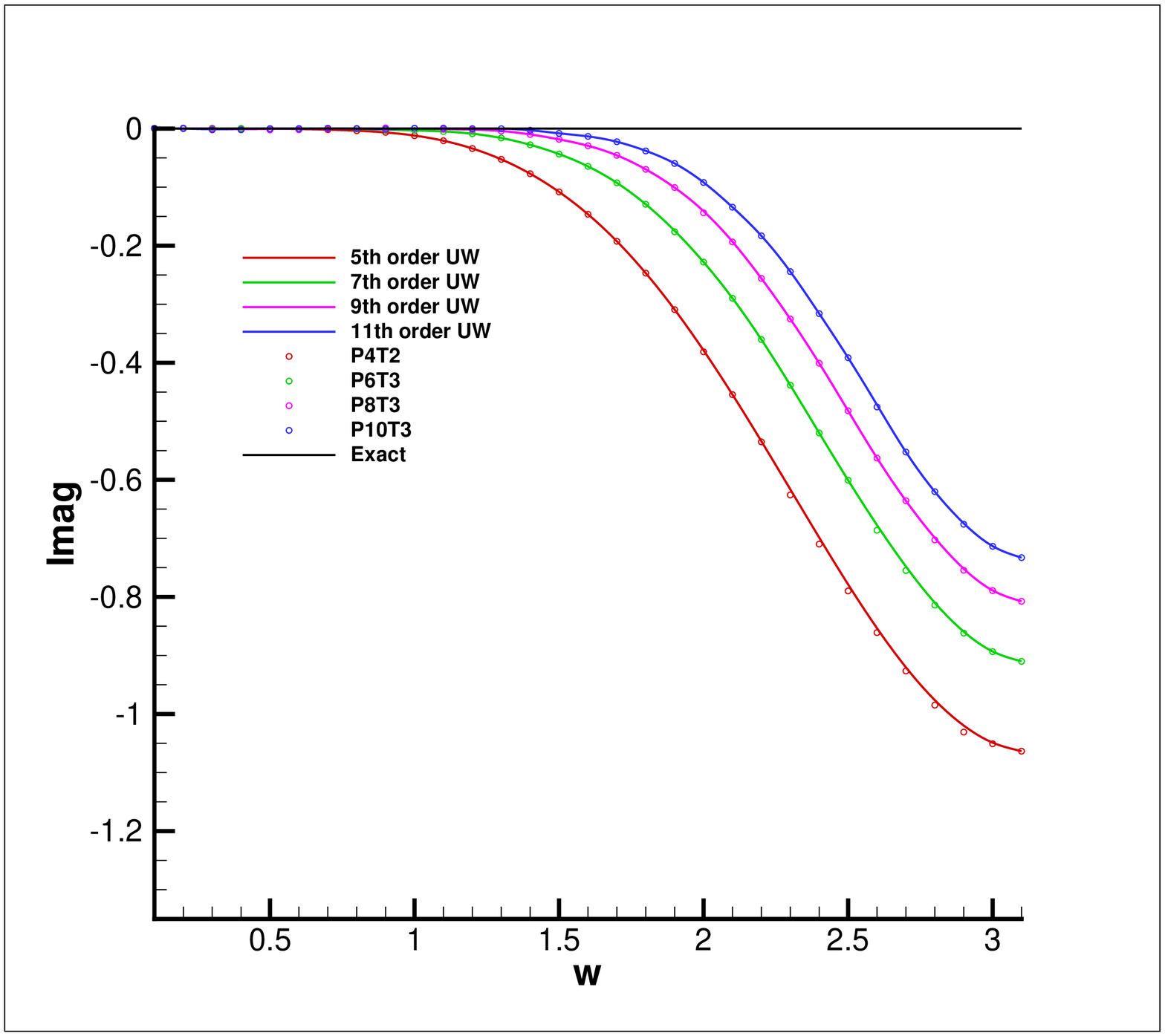}}
		\protect\caption{Approximate dispersion and dissipation properties for different order schemes. Real parts of modified wavenumber are shown in the left panel, while imaginary parts are shown in the right. The circles represent the spectral property of proposed $\mathrm{P}_{n}\mathrm{T}_{m}-\mathrm{BVD}$ schemes at different wave numbers. The solid lines represent the spectral property of corresponding high order upwind schemes in the same color. 
			\label{fig:ADR}}
	\end{center}	
\end{figure}

We study the spectral property of the proposed scheme by using approximate dispersion relation (ADR) analysis described in \cite{adr}. The numerical dissipation of a scheme can be quantified through the imaginary parts of the modified wavenumber while the numerical dispersion can be quantified with real parts. The spectral properties of different order schemes are shown in Fig.~\ref{fig:ADR} in which the spectral properties of proposed $\mathrm{P}_{n}\mathrm{T}_{m}-\mathrm{BVD}$ schemes at different wave numbers are marked by the circles, and the solid lines in the  same color represent the spectral properties of the corresponding high order-linear schemes.  It can be seen that the proposed schemes have almost the same spectral property as their underlying linear upwind schemes even at high wavenumber band. However, for WENO-type high order schemes, as shown in \cite{adr,TENO} accuracy is usually undermined at high wavenumber regime although they can recover to their underlying linear scheme at low wavenumber. The reason is that the WENO smoothness indicators tends to mis-interpret high frequency waves as discontinuities. Contrarily, the proposed $\mathrm{P}_{n}\mathrm{T}_{m}-\mathrm{BVD}$ schemes are designed to reduce numerical dissipations, and thus  preserve the high-order upwind schemes even in high wavenumber regime, which is hardly realized by any existing high resolution scheme based on limiting projections using nonlinear weighting.

\section{Numerical results \label{sec:results}}

In this section, some numerical experiments are performed as a demonstration of the proposed schemes. Linear advection equations and Euler equation systems will be numerically solved. In all examples, the ratio of specific heats is set to $\gamma=7/5$ and CFL is set to $0.4$.

\subsection{Accuracy test for advection of one-dimensional sine wave}\label{accuracy1}
In order to evaluate the convergence rate of the proposed $\mathrm{P}_{n}\mathrm{T}_{m}-\mathrm{BVD}$ schemes, an advection test of smooth profile was conducted on gradually refined grids. The initial smooth distribution was given by
\begin{equation}
q\left(x\right)=\sin\left(2\pi x\right), \ x\in\left[-1,1\right].
\end{equation}

We ran the computation for one period (at $t=2.0$) and summarized the numerical errors and the convergence rates for $\mathrm{P}_{n}\mathrm{T}_{m}-\mathrm{BVD}$ schemes in Table \ref{Tab:rate}. In order to compare with the corresponding high order upwind schemes, we also summarized numerical errors calculated by high order upwind schemes in Table \ref{Tab:rateUW}. As expected, the proposed $\mathrm{P}_{n}\mathrm{T}_{m}-\mathrm{BVD}$ schemes achieved $n+1$ order convergence rates when grid elements were gradually refined. Importantly, we observed that the $L_{1}$ and $L_{\infty}$ errors from the $\mathrm{P}_{n}\mathrm{T}_{m}-\mathrm{BVD}$ schemes were exactly same as those calculated by their corresponding high order linear upwind schemes, which was in line with the conclusion from the spectral property analysis in the previous section. These results confirm that the $\mathrm{P}_{n}\mathrm{T}_{m}-\mathrm{BVD}$ schemes are able to recover their underlying high order polynomial interpolations for smooth solutions.    

\begin{table}[]
	\centering
	\caption{Numerical errors and convergence rate for linear advection test. Results are computed by the proposed $\mathrm{P}_{n}\mathrm{T}_{m}-\mathrm{BVD}$ schemes.}
	\label{Tab:rate}
	\begin{tabular}{l|lllll}
		\hline
		Schemes                                      & Mesh & $L_{1}$ errors & $L_{1}$ order & $L_{\infty}$ errors & $L_{\infty}$ order \\ \hline
				\multirow{5}{*}{$\mathrm{P_{4}T_{2}}$}
& 10   & $2.493\times10^{-1}$ &  & $3.852\times10^{-1}$ & \\		                     
& 20   & $1.174\times10^{-2}$ & 4.41 & $1.815\times10^{-2}$ & 4.41\\
& 40   & $3.986\times10^{-4}$ & 4.88 & $6.309\times10^{-4}$ & 4.85 \\
& 80  & $1.274\times10^{-5}$ & 4.97 & $2.002\times10^{-5}$ & 4.98  \\ \hline
				\multirow{5}{*}{$\mathrm{P_{6}T_{3}}$}
& 10   & $8.518\times10^{-2}$ &  & $1.316\times10^{-1}$ & \\		                     
& 20   & $9.673\times10^{-4}$ & 6.46 & $1.495\times10^{-3}$ & 6.46\\
& 40   & $8.350\times10^{-6}$ & 6.86 & $1.319\times10^{-5}$ & 6.82 \\
& 80  & $6.686\times10^{-8}$ & 6.96 & $1.052\times10^{-7}$ & 6.97 \\ \hline
				\multirow{5}{*}{$\mathrm{P_{8}T_{3}}$}
& 10   & $2.733\times10^{-2}$ &  & $4.223\times10^{-2}$ & \\		                     
& 20   & $8.216\times10^{-5}$ & 8.38 & $1.269\times10^{-4}$ & 8.38\\
& 40   & $1.816\times10^{-7}$ & 8.82 & $2.870\times10^{-7}$ & 8.79 \\
& 80  & $3.659\times10^{-10}$ & 8.96 & $5.756\times10^{-10}$ & 8.96 \\ \hline
				\multirow{5}{*}{$\mathrm{P_{10}T_{3}}$}
& 10   & $8.716\times10^{-3}$ &  & $1.347\times10^{-2}$ & \\		                     
& 20   & $7.132\times10^{-6}$ & 10.26 & $1.102\times10^{-5}$ & 10.26\\
& 40   & $4.041\times10^{-9}$ & 10.79 & $6.388\times10^{-9}$ & 10.75 \\
& 80  & $2.051\times10^{-12}$ & 10.94 & $3.227\times10^{-12}$ & 10.95 \\ \hline
	\end{tabular}
\end{table}

\begin{table}[]
	\centering
	\caption{Same as table \ref{Tab:rate}, but by the high-order linear upwind schemes.}
	\label{Tab:rateUW}
	\begin{tabular}{l|lllll}
		\hline
		Schemes                                      & Mesh & $L_{1}$ errors & $L_{1}$ order & $L_{\infty}$ errors & $L_{\infty}$ order \\ \hline
		\multirow{5}{*}{5th order UW}
		& 10   & $2.493\times10^{-1}$ &  & $3.852\times10^{-1}$ & \\		                     
		& 20   & $1.174\times10^{-2}$ & 4.41 & $1.815\times10^{-2}$ & 4.41\\
		& 40   & $3.986\times10^{-4}$ & 4.88 & $6.309\times10^{-4}$ & 4.85 \\
		& 80  & $1.274\times10^{-5}$ & 4.97 & $2.002\times10^{-5}$ & 4.98  \\ \hline
		\multirow{5}{*}{7th order UW}
		& 10   & $8.518\times10^{-2}$ &  & $1.316\times10^{-1}$ & \\		                     
		& 20   & $9.673\times10^{-4}$ & 6.46 & $1.495\times10^{-3}$ & 6.46\\
		& 40   & $8.350\times10^{-6}$ & 6.86 & $1.319\times10^{-5}$ & 6.82 \\
		& 80  & $6.686\times10^{-8}$ & 6.96 & $1.052\times10^{-7}$ & 6.97 \\ \hline
		\multirow{5}{*}{9th order UW}
		& 10   & $2.733\times10^{-2}$ &  & $4.223\times10^{-2}$ & \\		                     
		& 20   & $8.216\times10^{-5}$ & 8.38 & $1.269\times10^{-4}$ & 8.38\\
		& 40   & $1.816\times10^{-7}$ & 8.82 & $2.870\times10^{-7}$ & 8.79 \\
		& 80  & $3.659\times10^{-10}$ & 8.96 & $5.756\times10^{-10}$ & 8.96 \\ \hline
		\multirow{5}{*}{11th order UW}
		& 10   & $8.716\times10^{-3}$ &  & $1.347\times10^{-2}$ & \\		                     
		& 20   & $7.132\times10^{-6}$ & 10.26 & $1.102\times10^{-5}$ & 10.26\\
		& 40   & $4.041\times10^{-9}$ & 10.79 & $6.388\times10^{-9}$ & 10.75 \\
		& 80  & $2.051\times10^{-12}$ & 10.94 & $3.227\times10^{-12}$ & 10.95 \\ \hline
	\end{tabular}
\end{table}

\subsection{Accuracy test for advection of a smooth profile containing critical points}\label{accuracy2}
We conducted the accuracy test which was more challenging for numerical schemes to distinguish smooth and non-smooth profiles because the initial distribution contains critical points. It has been reported in \cite{WENOM} that WENO type schemes do not reach their formal order of accuracy at critical points where the high order derivative does not simultaneously vanish. Following \cite{WENOM,veryhigh}, the initial condition was given by
\begin{equation}
q\left(x\right)=\sin(\pi x-\dfrac{\sin(\pi x)}{\pi}), \ x\in\left[-1,1\right].
\end{equation}
The computation was conducted for ten period ($t=20$). We summarize the numerical errors $L_{1}$ and $L_{\infty}$ of the proposed $\mathrm{P}_{n}\mathrm{T}_{m}-\mathrm{BVD}$ in Table \ref{Tab:critical} and those of the corresponding upwind schemes in Table \ref{Tab:criticalUW}. It can be seen that the $\mathrm{P}_{n}\mathrm{T}_{m}-\mathrm{BVD}$ schemes are capable of achieving their highest possible order of accuracy in the grid refinement tests. Compared with their corresponding upwind schemes, the $\mathrm{P}_{n}\mathrm{T}_{m}-\mathrm{BVD}$ schemes almost realize the same $L_{1}$ and $L_{\infty}$ errors, which demonstrates that $\mathrm{P}_{n}\mathrm{T}_{m}-\mathrm{BVD}$ schemes restore upwind schemes even at critical points. However, as shown in \cite{WENOM,veryhigh}, it is generally difficult for WENO type schemes to recover their underlying high order upwind schemes particularly around the critical points. 

\begin{table}[]
	\centering
	\caption{Numerical errors and convergence rate for advection of a smooth profile containing critical points. Results are calculated by $\mathrm{P}_{n}\mathrm{T}_{m}-\mathrm{BVD}$ schemes.}
	\label{Tab:critical}
	\begin{tabular}{l|lllll}
		\hline
		Schemes                                      & Mesh & $L_{1}$ errors & $L_{1}$ order & $L_{\infty}$ errors & $L_{\infty}$ order \\ \hline
		\multirow{5}{*}{$\mathrm{P_{4}T_{2}}$}
		& 10   & $1.068\times10^{-1}$ &  & $2.263\times10^{-1}$ & \\		                     
		& 20   & $1.497\times10^{-2}$ & 2.83 & $3.772\times10^{-2}$ & 2.58\\
		& 40   & $7.138\times10^{-4}$ & 5.13 & $1.960\times10^{-3}$ & 4.27 \\
		& 80  & $2.327\times10^{-5}$ & 4.94 & $6.582\times10^{-5}$ & 4.90  \\
		& 160  & $7.334\times10^{-7}$ & 4.99 & $2.092\times10^{-6}$ & 4.98  \\ \hline
		\multirow{5}{*}{$\mathrm{P_{6}T_{3}}$}
		& 10   & $7.969\times10^{-2}$ &  & $1.306\times10^{-1}$ & \\		                     
		& 20   & $2.539\times10^{-3}$ & 4.97 & $6.565\times10^{-3}$ & 4.31\\
		& 40   & $2.610\times10^{-5}$ & 6.60 & $8.020\times10^{-5}$ & 6.36 \\
		& 80  & $2.135\times10^{-7}$ & 6.93 & $6.654\times10^{-7}$ & 6.91 \\
		& 160  & $1.691\times10^{-9}$ & 6.98 & $5.269\times10^{-9}$ & 6.98 \\ \hline
		\multirow{5}{*}{$\mathrm{P_{8}T_{3}}$}
		& 10   & $3.702\times10^{-2}$ &  & $6.049\times10^{-2}$ & \\		                     
		& 20   & $5.008\times10^{-4}$ & 6.21 & $1.287\times10^{-3}$ & 5.55\\
		& 40   & $1.413\times10^{-6}$ & 8.47 & $4.492\times10^{-6}$ & 8.16 \\
		& 80  & $2.975\times10^{-9}$ & 8.89 & $9.779\times10^{-9}$ & 8.84 \\
		& 160  & $5.942\times10^{-12}$ & 8.97 & $1.960\times10^{-11}$ & 8.96 \\ \hline
		\multirow{5}{*}{$\mathrm{P_{10}T_{3}}$}
		& 10   & $1.449\times10^{-2}$ &  & $2.837\times10^{-2}$ & \\		                     
		& 20   & $1.222\times10^{-4}$ & 6.89 & $3.180\times10^{-4}$ & 6.48\\
		& 40   & $9.755\times10^{-8}$ & 10.29 & $3.357\times10^{-7}$ & 9.89 \\
		& 80  & $5.472\times10^{-11}$ & 10.80 & $1.919\times10^{-10}$ & 10.77 \\
		& 160  & $4.032\times10^{-14}$ & 10.40 & $1.472\times10^{-13}$ & 10.35 \\ \hline
	\end{tabular}
\end{table}

\begin{table}[]
	\centering
	\caption{Same as table \ref{Tab:critical}, but by high-order linear upwind schemes.}
	\label{Tab:criticalUW}
	\begin{tabular}{l|lllll}
		\hline
		Schemes                                      & Mesh & $L_{1}$ errors & $L_{1}$ order & $L_{\infty}$ errors & $L_{\infty}$ order \\ \hline
			\multirow{5}{*}{5th order UW}
	& 10   & $1.067\times10^{-1}$ &  & $2.265\times10^{-1}$ & \\		                     
	& 20   & $1.793\times10^{-2}$ & 2.57 & $3.874\times10^{-2}$ & 2.55\\
	& 40   & $7.138\times10^{-4}$ & 4.65 & $1.960\times10^{-3}$ & 4.30 \\
   & 80  & $2.327\times10^{-5}$ & 4.94 & $6.582\times10^{-5}$ & 4.90  \\
    & 160  & $7.334\times10^{-7}$ & 4.99 & $2.092\times10^{-6}$ & 4.98  \\ \hline
		\multirow{5}{*}{7th order UW}
	& 10   & $7.969\times10^{-2}$ &  & $1.306\times10^{-1}$ & \\		                     
& 20   & $2.539\times10^{-3}$ & 4.97 & $6.565\times10^{-3}$ & 4.31\\
& 40   & $2.610\times10^{-5}$ & 6.60 & $8.020\times10^{-5}$ & 6.36 \\
& 80  & $2.135\times10^{-7}$ & 6.93 & $6.654\times10^{-7}$ & 6.91 \\
& 160  & $1.691\times10^{-9}$ & 6.98 & $5.269\times10^{-9}$ & 6.98 \\ \hline
		\multirow{5}{*}{9th order UW}
		& 10   & $3.702\times10^{-2}$ &  & $6.048\times10^{-2}$ & \\		                     
		& 20   & $5.008\times10^{-4}$ & 6.21 & $1.287\times10^{-3}$ & 5.55\\
& 40   & $1.413\times10^{-6}$ & 8.47 & $4.492\times10^{-6}$ & 8.16 \\
& 80  & $2.975\times10^{-9}$ & 8.89 & $9.779\times10^{-9}$ & 8.84 \\
& 160  & $5.942\times10^{-12}$ & 8.97 & $1.960\times10^{-11}$ & 8.96 \\ \hline
		\multirow{5}{*}{11th order UW}
		& 10   & $1.449\times10^{-2}$ &  & $2.837\times10^{-2}$ & \\		                     
& 20   & $1.222\times10^{-4}$ & 6.89 & $3.180\times10^{-4}$ & 6.48\\
& 40   & $9.755\times10^{-8}$ & 10.29 & $3.357\times10^{-7}$ & 9.89 \\
& 80  & $5.472\times10^{-11}$ & 10.80 & $1.919\times10^{-10}$ & 10.77 \\
& 160  & $4.032\times10^{-14}$ & 10.40 & $1.472\times10^{-13}$ & 10.35 \\ \hline
	\end{tabular}
\end{table}

\subsection{Advection of complex waves}
\begin{figure}
	\begin{center}
   \subfigure[$\mathrm{P_{4}T_{2}}$]{\includegraphics[scale=0.35,trim={0.9cm 0.9cm 0.9cm 0.9cm},clip]{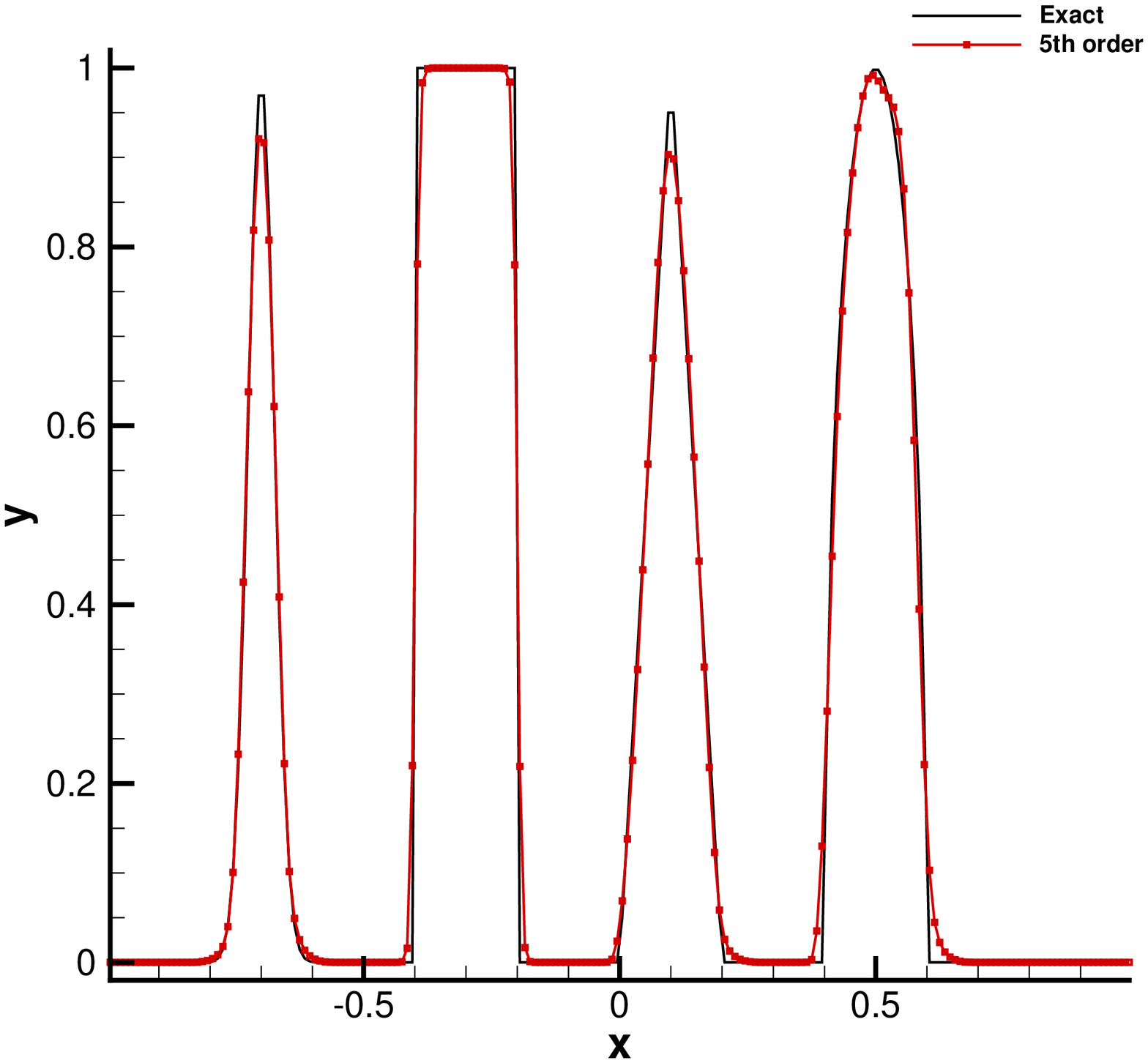}}
   \subfigure[$\mathrm{P_{6}T_{3}}$]{\includegraphics[scale=0.35,trim={0.9cm 0.9cm 0.9cm 0.9cm},clip]{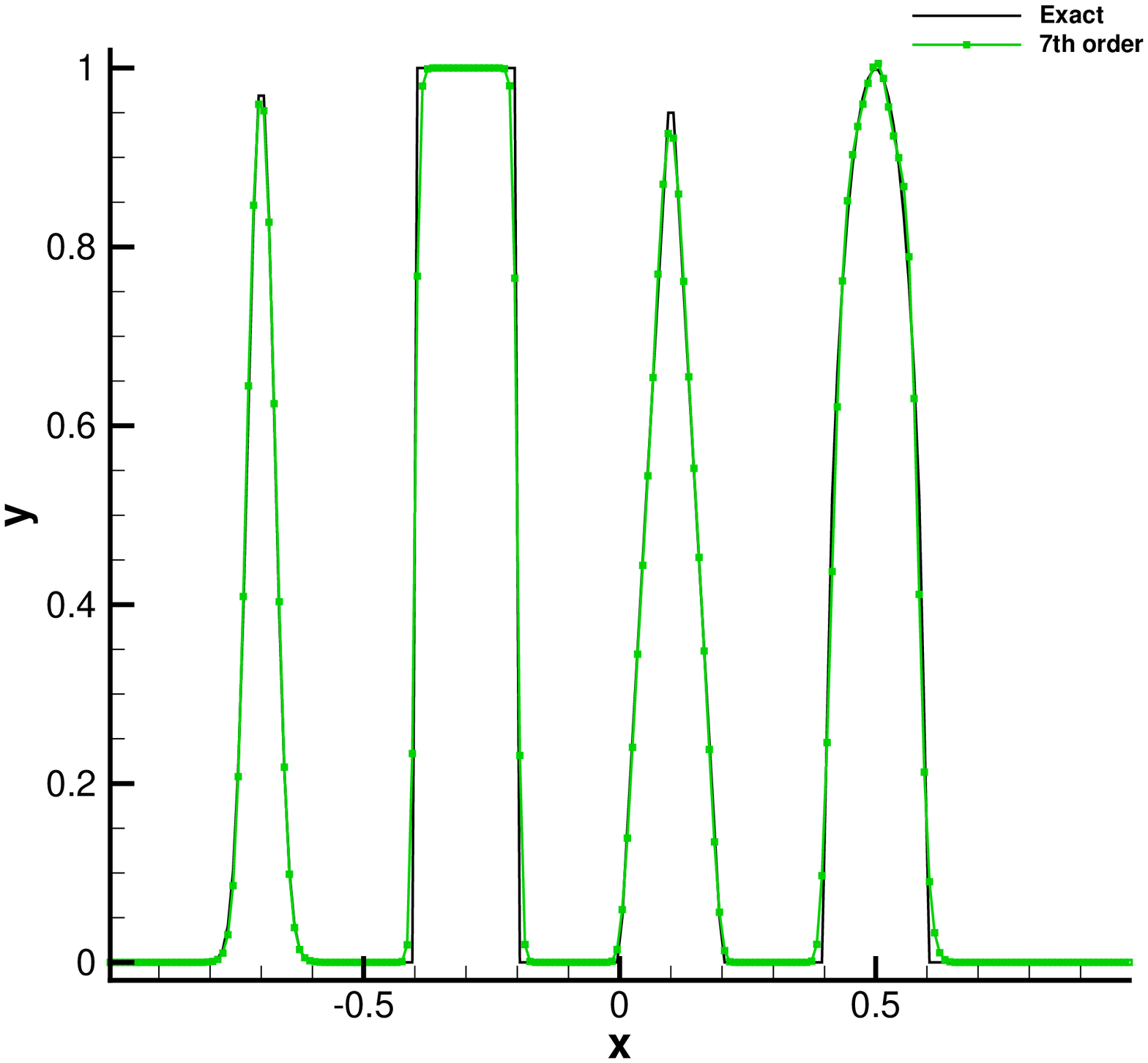}}
   \subfigure[$\mathrm{P_{8}T_{3}}$]{\includegraphics[scale=0.35,trim={0.9cm 0.9cm 0.9cm 0.9cm},clip]{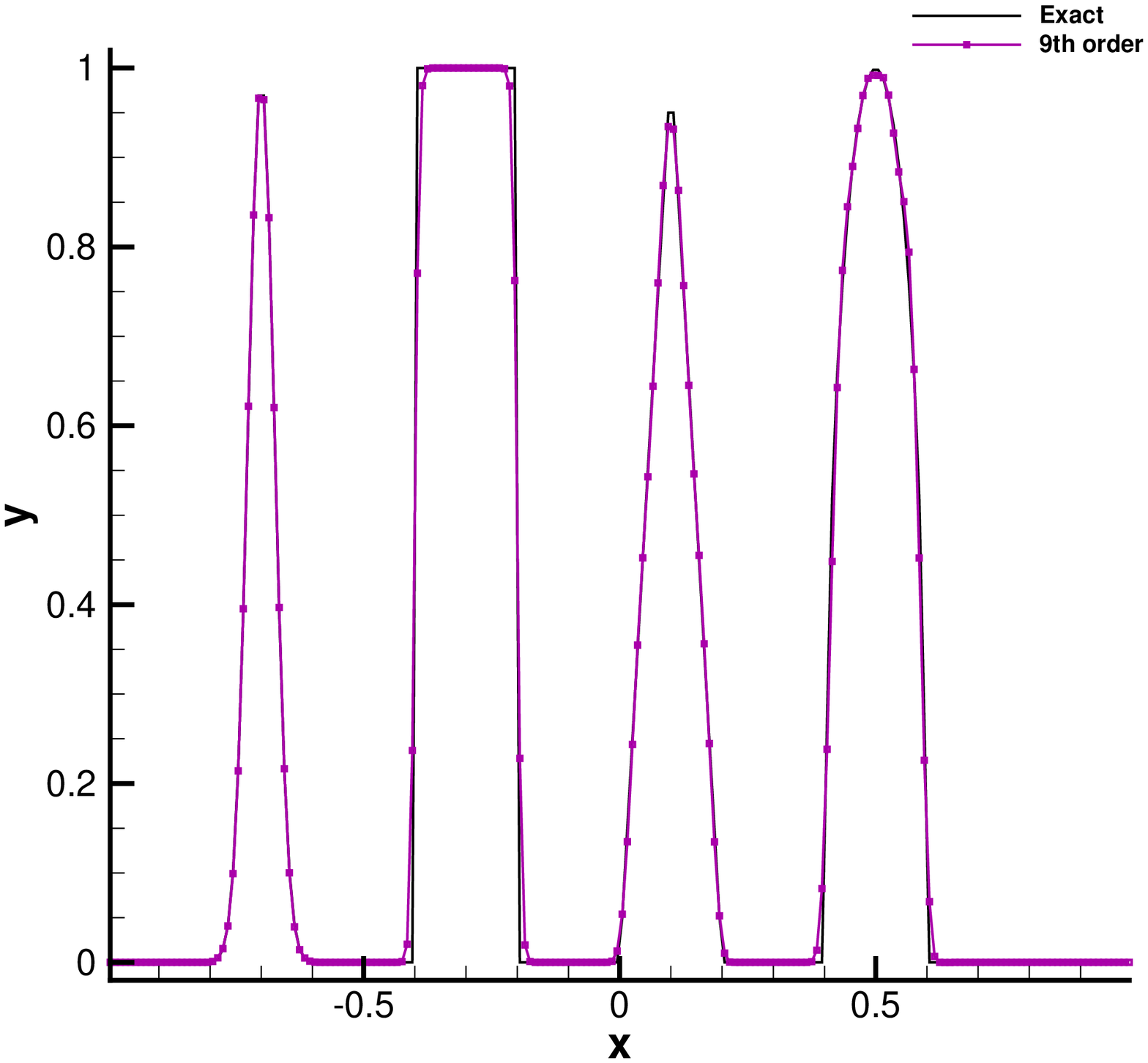}}
  \subfigure[$\mathrm{P_{10}T_{3}}$]{\includegraphics[scale=0.35,trim={0.9cm 0.9cm 0.9cm 0.9cm},clip]{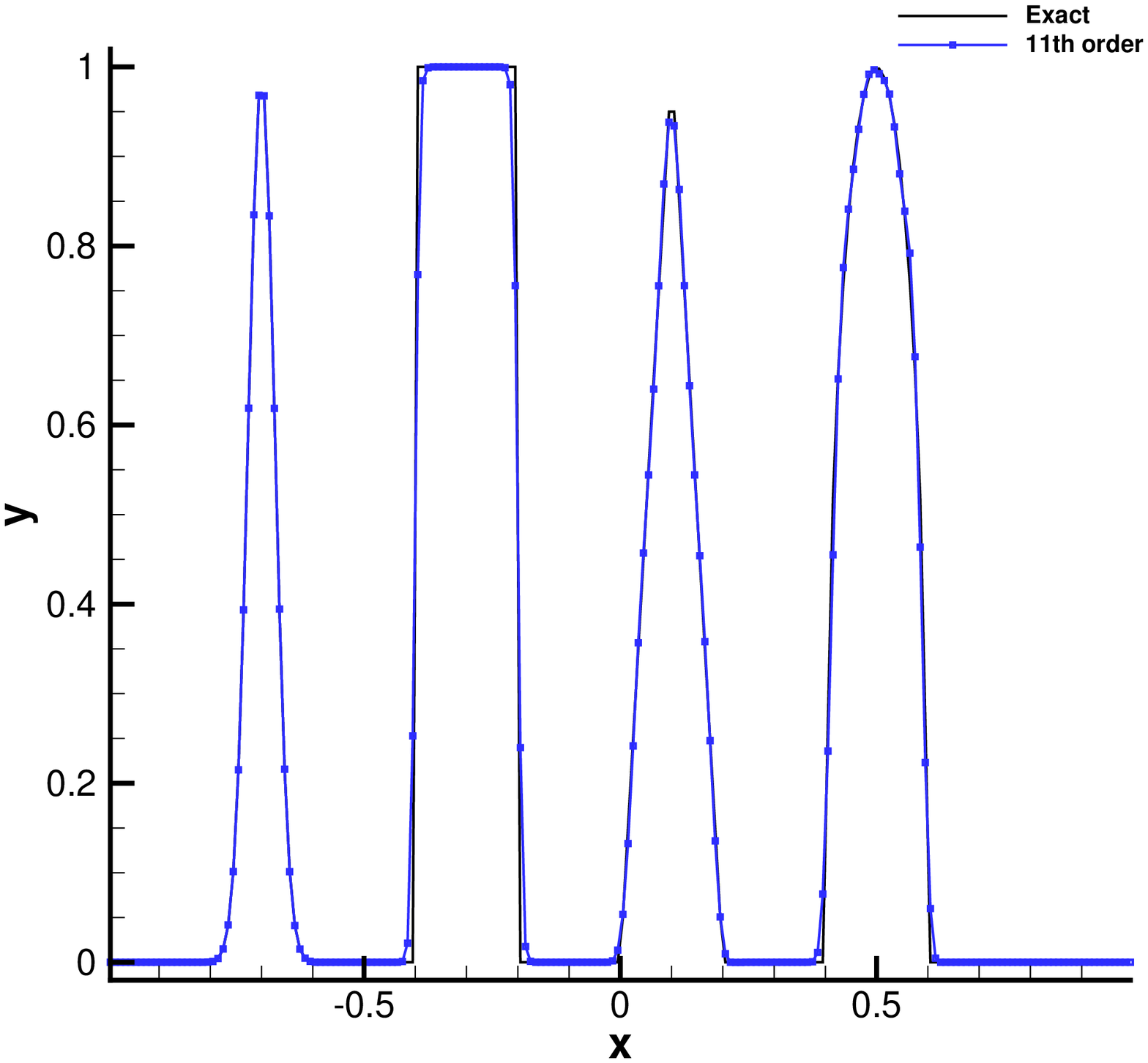}}
  \end{center}
	\protect\caption{Numerical results for advection of complex waves. The numerical solutions at $t=2.0$ with 200 mesh cells are presented.
		\label{fig:shujiang}}	
\end{figure}

In order to examine the performance of the proposed scheme in solving profiles of different smoothness, we further simulated the propagation of a complex wave \cite{jiang96}. The initial profile contains both discontinuities and smooth regions with different smoothness. The computation was carried out for one period at $t=2.0$ with a 200-cell mesh. The results calculated by the $\mathrm{P}_{n}\mathrm{T}_{m}-\mathrm{BVD}$ schemes were presented in Fig.~\ref{fig:shujiang}. It can be seen that all of schemes are free of visible numerical oscillations and capture sharper discontinuities than conventional high order FVM schemes. With polynomial degree increased, the extreme points of the initial profile are better resolved. It is also noted that $\mathrm{P}_{n}\mathrm{T}_{m}-\mathrm{BVD}$ schemes produce almost same results for the discontinuity which is resolved by only four cells. Because the BVD algorithm can properly choose THINC reconstruction function across discontinuities, increasing order is only effective to the resolution of smooth regions.   

\subsection{Sod's problem}
\begin{figure}
	\begin{center}
	\subfigure[$\mathrm{P_{4}T_{2}}$]{\includegraphics[scale=0.35,trim={0.5cm 0.5cm 0.5cm 0.5cm},clip]{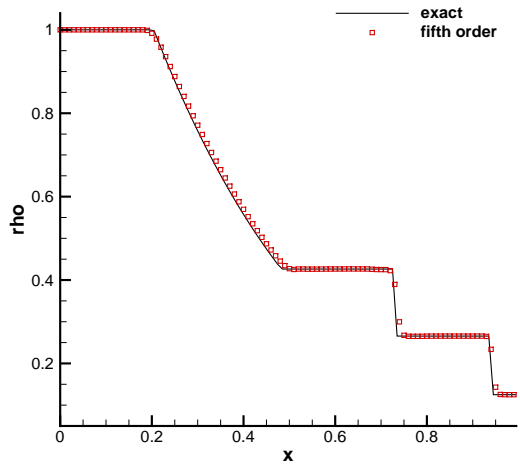}}
	\subfigure[$\mathrm{P_{6}T_{3}}$]{\includegraphics[scale=0.35,trim={0.5cm 0.5cm 0.5cm 0.5cm},clip]{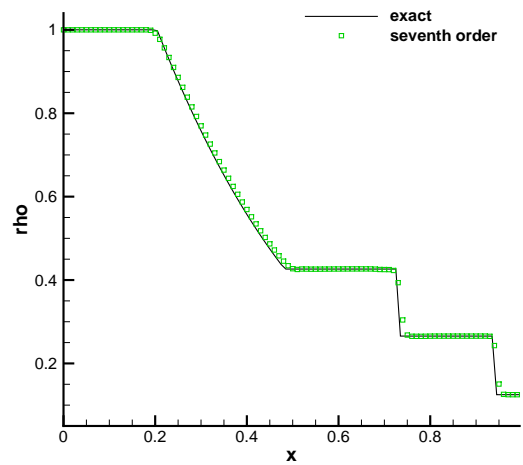}}
	\subfigure[$\mathrm{P_{8}T_{3}}$]{\includegraphics[scale=0.35,trim={0.5cm 0.5cm 0.5cm 0.5cm},clip]{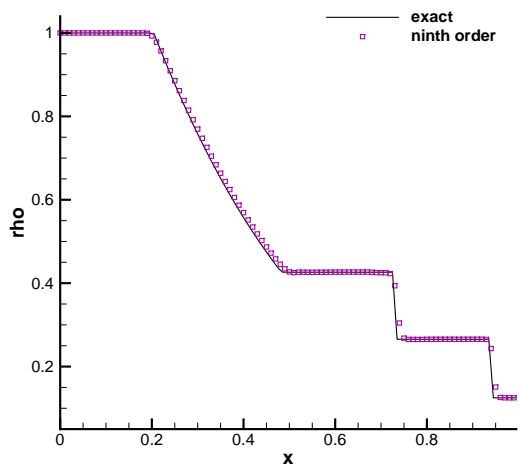}}
	\subfigure[$\mathrm{P_{10}T_{3}}$]{\includegraphics[scale=0.35,trim={0.5cm 0.5cm 0.5cm 0.5cm},clip]{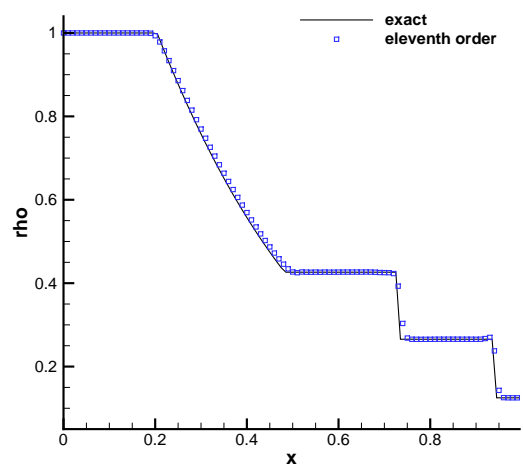}}
	\end{center}
	\protect\caption{Numerical results of Sod's problem for density field at $t = 0.25$ with $100$ cells. 
    \label{fig:sodrho}}	
\end{figure}

As one of widely used benchmark tests for shock-capturing schemes, the Sod's problem was employed to test the performance of present schemes in capturing the shock front,  contact discontinuity, as well as the expansion wave. The initial distribution on computational domain $[0,1]$ was specified as \cite{sod}
\begin{equation}
\left(\rho_{0},\ u_{0},\ p_{0}\right)=\left\{
\begin{array}{ll}
\left(1,\ 0,\ 1\right) & 0 \leq x \leq 0.5 \\
\left(0.125,\ 0,\ 0.1\right) & \mathrm{otherwise}
\end{array}
\right..
\end{equation}
The computation was carried out on a mesh of  100 uniform cells up to $t=0.25$. The numerical results calculated from the proposed scheme were shown in Fig.~\ref{fig:sodrho} for density fields. From the results, we observe that $\mathrm{P}_{n}\mathrm{T}_{m}-\mathrm{BVD}$ schemes can solve the contact discontinuity without obvious numerical oscillations. Compared with the results produced by other high order shock-capturing schemes \cite{jiang96, WENOM, wenoz, wenoh, TENO,embedded}, the results of the proposed schemes are among the best with the jump contact resolved within only two cells. We also observe that $\mathrm{P}_{n}\mathrm{T}_{m}-\mathrm{BVD}$ schemes of different orders produce similar results across the contact where the THINC function is selected by the BVD algorithm. 

\subsection{Lax's problem}
\begin{figure}
	\begin{center}
	\subfigure[$\mathrm{P_{4}T_{2}}$]{\includegraphics[scale=0.35,trim={0.5cm 0.5cm 0.5cm 0.5cm},clip]{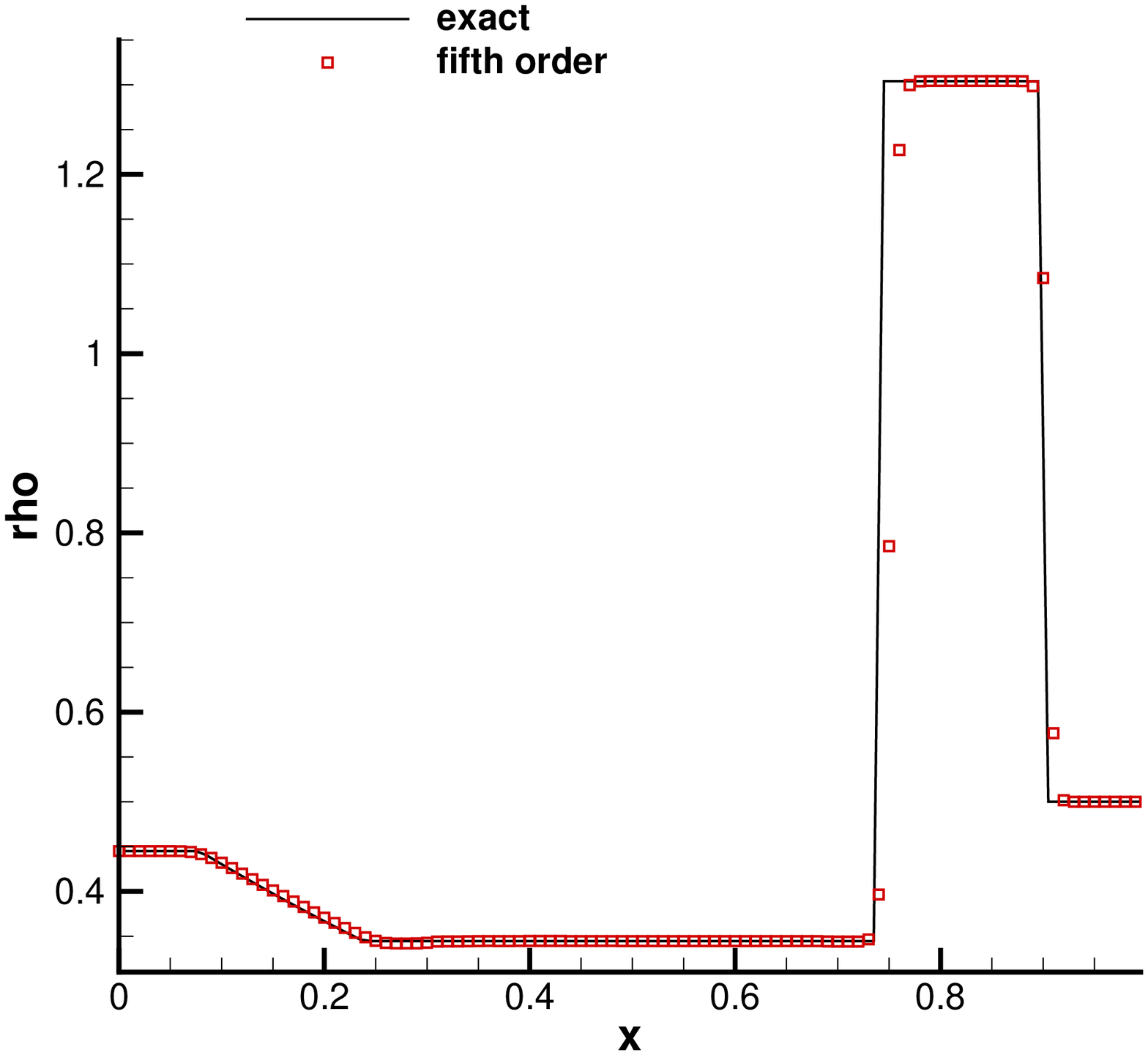}}
\subfigure[$\mathrm{P_{6}T_{3}}$]{\includegraphics[scale=0.35,trim={0.5cm 0.5cm 0.5cm 0.5cm},clip]{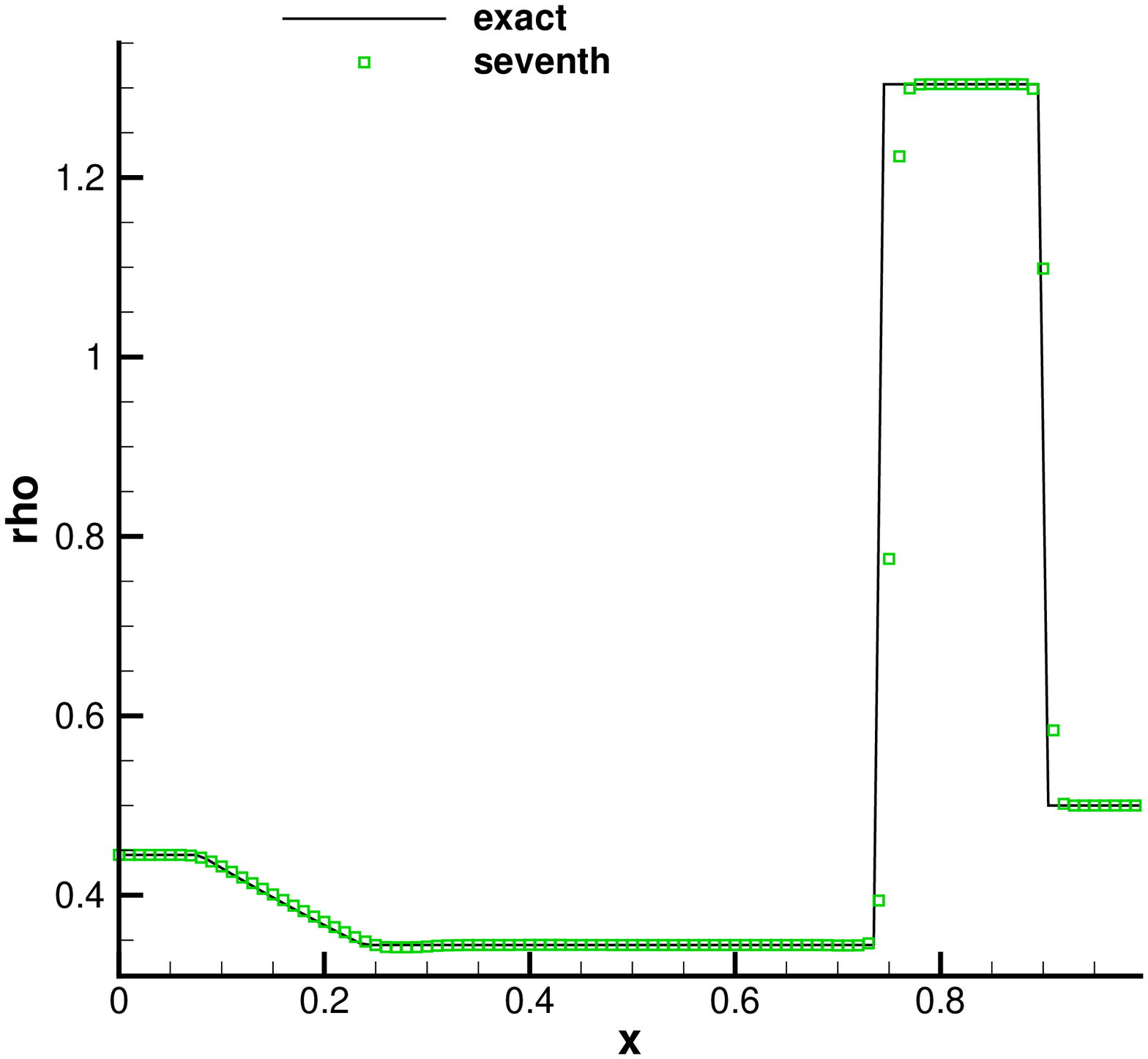}}
\subfigure[$\mathrm{P_{8}T_{3}}$]{\includegraphics[scale=0.35,trim={0.5cm 0.5cm 0.5cm 0.5cm},clip]{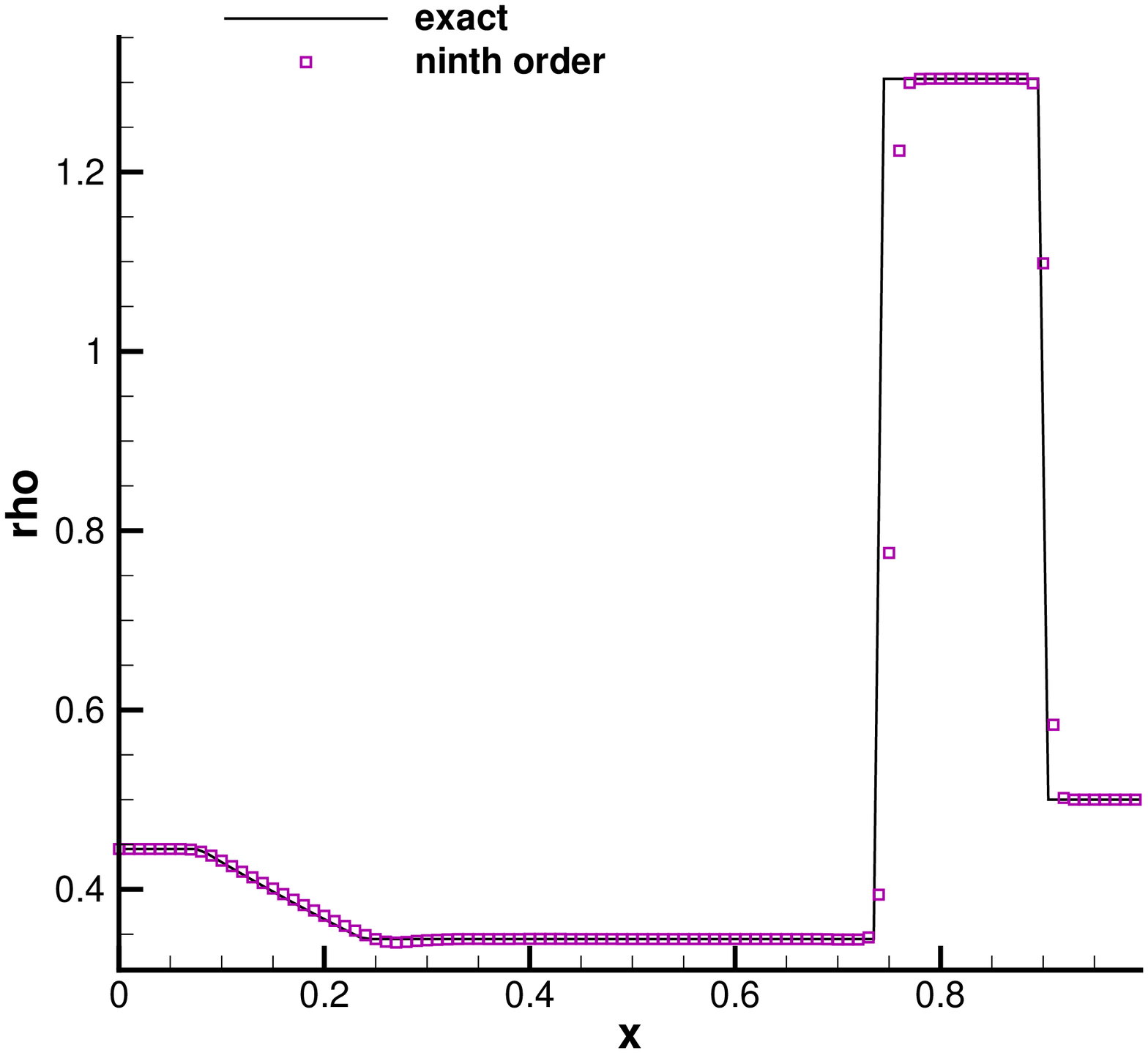}}
\subfigure[$\mathrm{P_{10}T_{3}}$]{\includegraphics[scale=0.35,trim={0.5cm 0.5cm 0.5cm 0.5cm},clip]{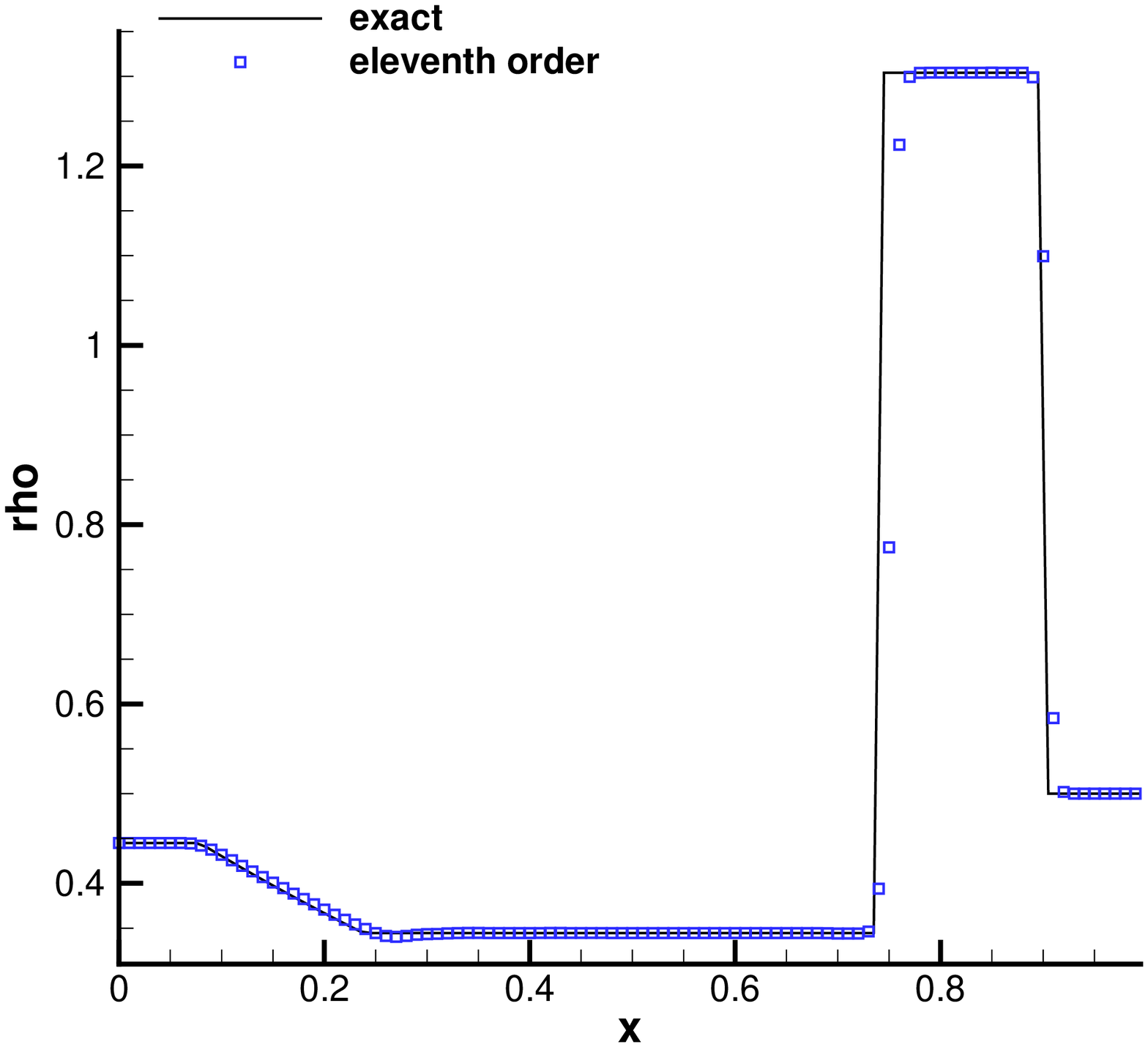}}
  \end{center}
  \protect\caption{Numerical results of Lax's problem for density field at time $t = 0.16$ with $100$ cells. 
    \label{fig:lax}}	
\end{figure}

To check the ability of the proposed numerical scheme 
to capture relatively strong shock, we solved the Lax problem \cite{shu_eno1} in this subsection. The initial condition is given by
\begin{equation}
\left(\rho_{0},\ u_{0},\ p_{0}\right)=\left\{
\begin{array}{ll}
\left(0.445,\ 0.698,\ 3.528\right) & 0 \leq x \leq0.5\\
\left(0.5,\ 0.0,\ 0.571\right) & \mathrm{otherwise}
\end{array}
\right..
\end{equation}
With the same number of cells as in the previous test case, we got the numerical results at $t=0.16$. The density field is plotted presented in Fig.~\ref{fig:lax}. Obviously, the proposed $\mathrm{P}_{n}\mathrm{T}_{m}-\mathrm{BVD}$ schemes obtain accurate solutions without numerical oscillations. Again, the BVD algorithm chooses the THINC function across the discontinuity, which produces similar results among $\mathrm{P}_{n}\mathrm{T}_{m}-\mathrm{BVD}$ schemes of different order. Compared with published works using WENO type schemes, the present results are one of best. 

\subsection{Shock density wave interaction problem}
\begin{figure}
	\subfigure[$\mathrm{P_{4}T_{2}}$ and $\mathrm{P_{6}T_{3}}$]
	{\centering\includegraphics[scale=0.35,trim={0.5cm 0.5cm 0.5cm 0.5cm},clip]{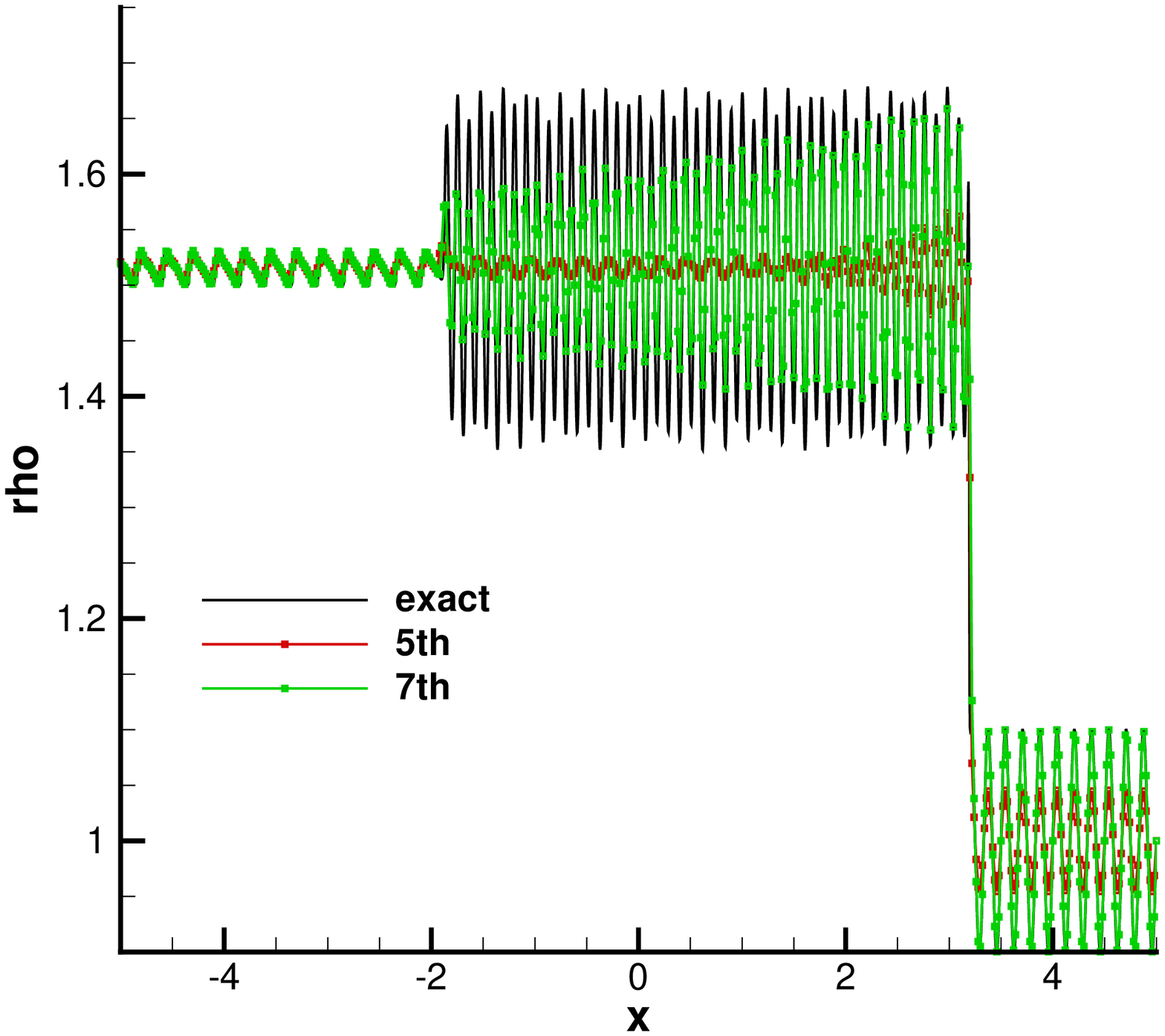}}
    \subfigure[$\mathrm{P_{8}T_{3}}$ and $\mathrm{P_{10}T_{3}}$]
	{\centering\includegraphics[scale=0.35,trim={0.5cm 0.5cm 0.5cm 0.5cm},clip]{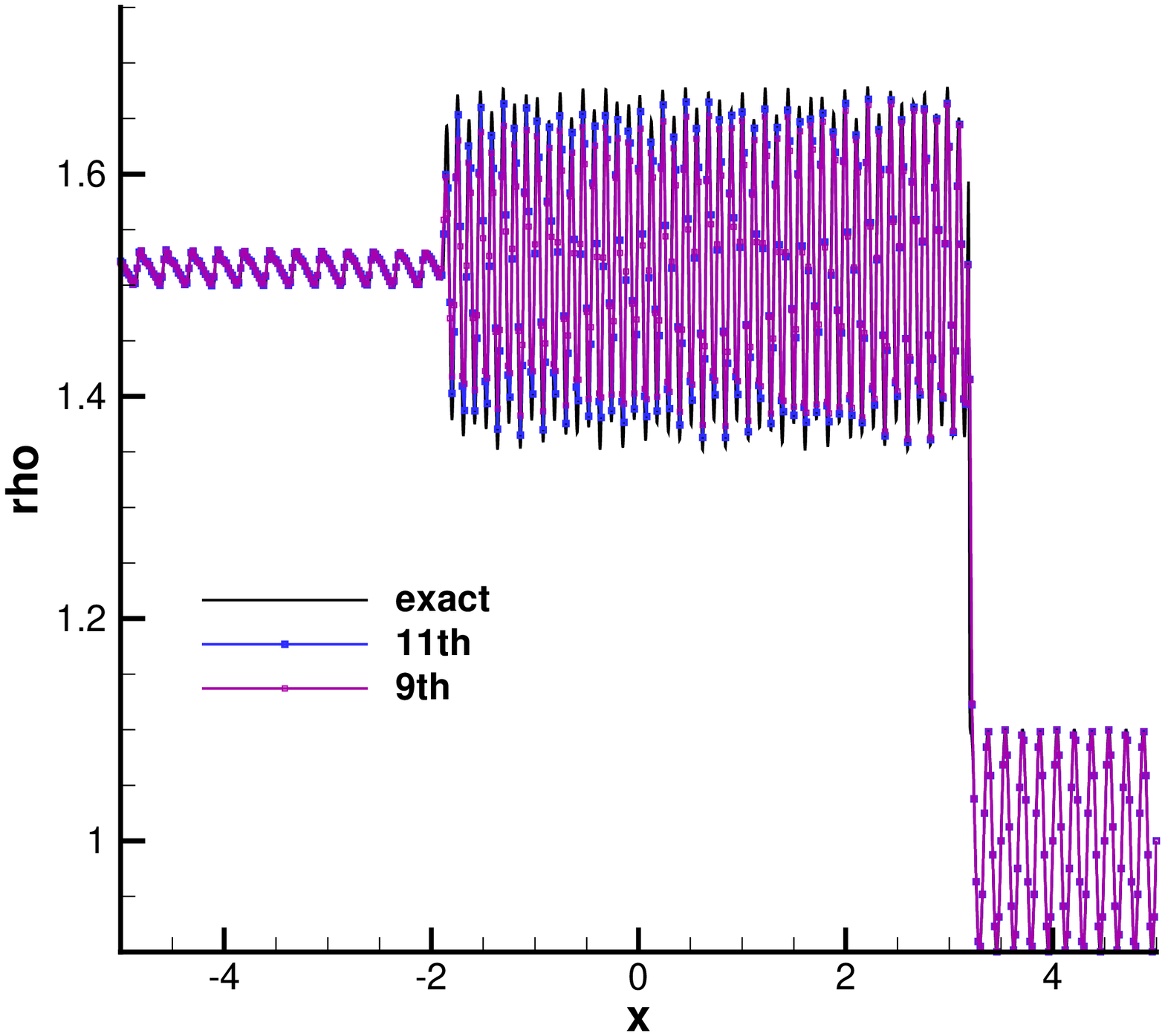}}
	\protect\caption{Numerical results of shock density wave interaction problem. The results of $\mathrm{P_{4}T_{2}}$ and $\mathrm{P_{6}T_{3}}$ are shown in the left panel, and $\mathrm{P_{8}T_{3}}$ and $\mathrm{P_{10}T_{3}}$ in the right panel.
		\label{fig:TT}}	
\end{figure}

\begin{figure}
	\subfigure[$\mathrm{P_{4}T_{2}}$ and $\mathrm{P_{6}T_{3}}$]
{\centering\includegraphics[scale=0.35,trim={0.5cm 0.5cm 0.5cm 0.5cm},clip]{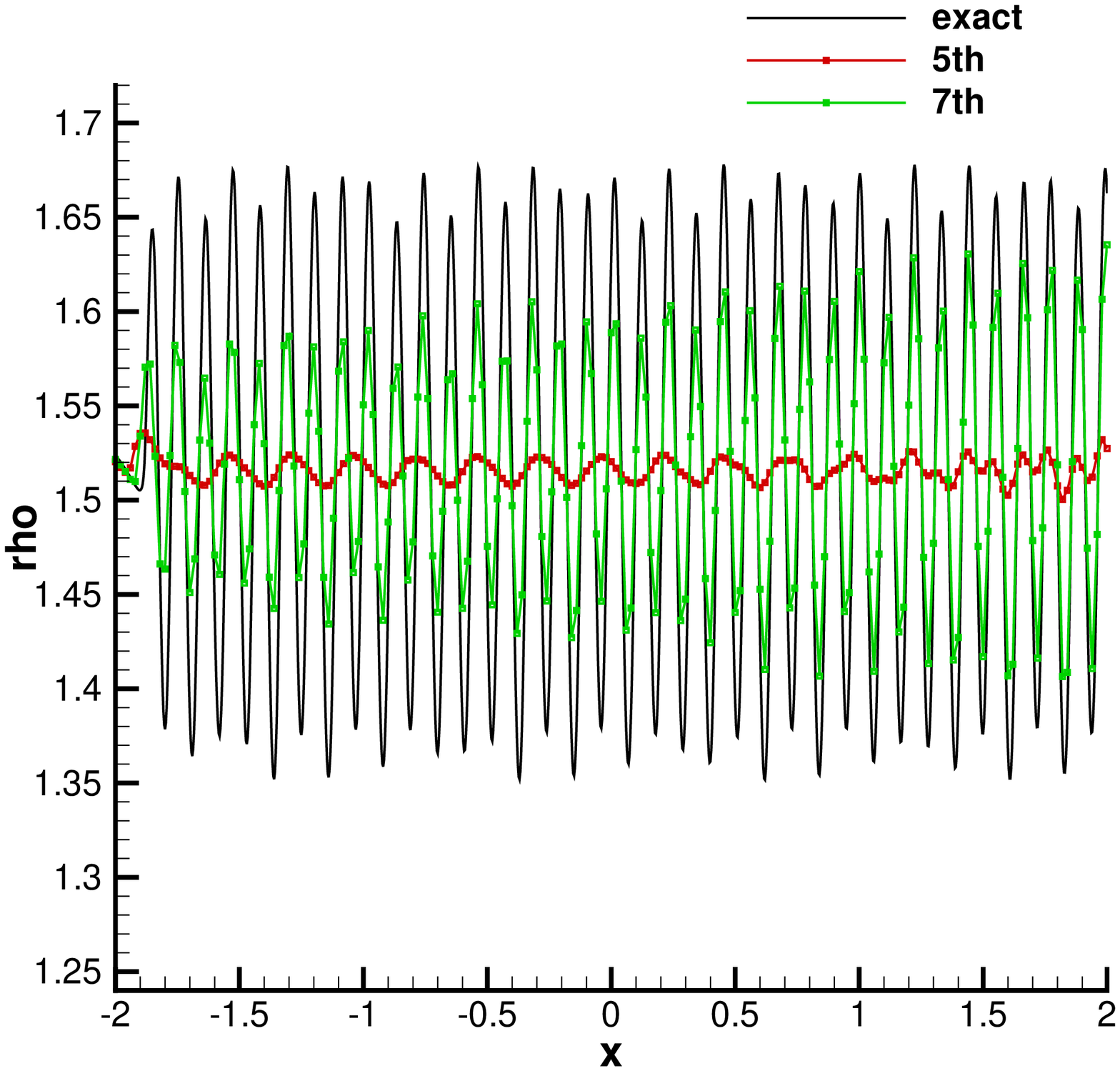}}
\subfigure[$\mathrm{P_{8}T_{3}}$ and $\mathrm{P_{10}T_{3}}$]
{\centering\includegraphics[scale=0.35,trim={0.5cm 0.5cm 0.5cm 0.5cm},clip]{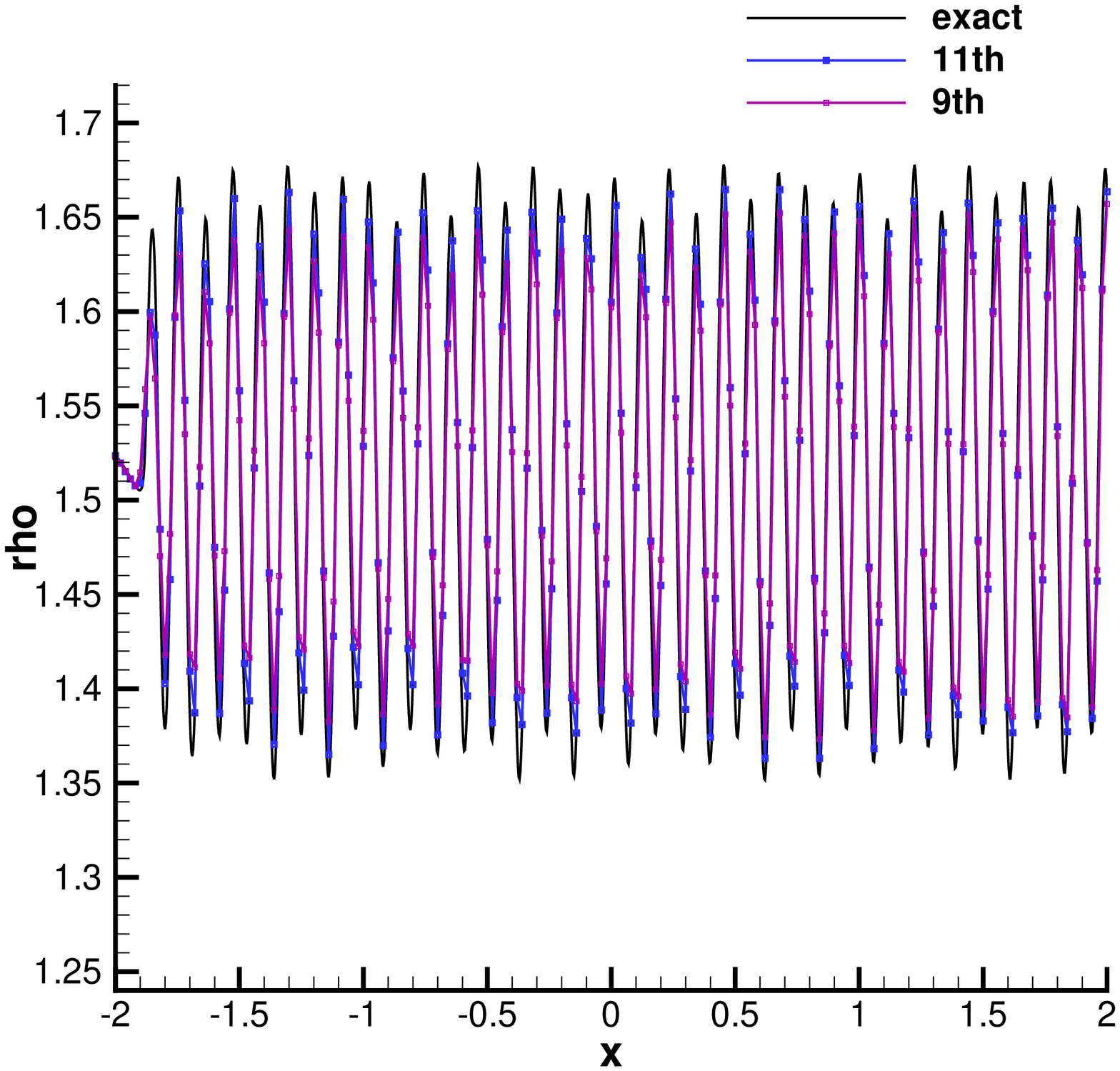}}
	\protect\caption{The same as the Fig.~\ref{fig:TT}, but showing a zoomed region of high-frequency waves.
		\label{fig:TT_L}}	
\end{figure}

In order to verify the performance of the present schemes in capturing shocks and smooth solutions of different scales, we simulated the case proposed in \cite{TitarevToro}, which serves as a good test bed for the simulations of compressible turbulence involving shock waves. In this case, a shock wave interacts with density disturbances and generates a flow field containing waves of higher wavenumber and discontinuities. The initial condition is specified similarly as \cite{TitarevToro}
\begin{equation}
(\rho_{0},\ u_{0},\ p_{0})=\left\{
\begin{array}{ll}
\left(1.515695,\ 0.523346,\ 1.805\right), \ &\mathrm{if}\  x \leq -4.5,\\
\left(1+0.1\sin(12 \pi x),\ 0,\ 1\right), \ &\mathrm{otherwise}.
\end{array}\right.
\end{equation}
The computation was carried out up to $t=5.0$. The numerical solutions with 500 cells were shown in Fig.~\ref{fig:TT}, where comparisons of different order schemes are  included. For better illustration, we also plot a zoomed region for density perturbation in Fig.~\ref{fig:TT_L}. In contrast to the Sod's and Lax's problems where similar results are  obtained by $\mathrm{P}_{n}\mathrm{T}_{m}-\mathrm{BVD}$ schemes of different orders, in this test case the resolution of high-frequency waves is significantly improved as the degree of polynomial increases. As shown in ADR analysis and accuracy tests in subsection \ref{accuracy1} and \ref{accuracy2}, BVD algorithm selects high order upwind schemes in smooth region. Thus the resolution of high-frequency waves is improved with the order of the underlying polynomial increased. This test confirms that $\mathrm{P}_{n}\mathrm{T}_{m}-\mathrm{BVD}$ schemes can simultaneously resolve flows containing both smooth and discontinuous regions with high accuracy.    

\subsection{Two interacting blast waves}
Interactive blast waves involving multiple interactions of strong shocks and rarefaction waves has been introduced in \cite{blast}. The initial distribution is given by
\begin{equation}
(\rho_{0},\ u_{0},\ p_{0})=\left\{
\begin{array}{lll}
\left(1,\ 0,\ 1000\right), \ &\mathrm{if}\ 0 \leq x <0.1,\\
\left(1,\ 0,\ 0.01\right), \ &\mathrm{if}\ 0.1 \leq x <0.9, \\
\left(1,\ 0,\ 100\right),  \ &\mathrm{if}\ 0.9 \leq x <1.
\end{array}\right.
\end{equation}

\begin{figure}
	\begin{center}
    \subfigure[$\mathrm{P_{4}T_{2}}$]{\includegraphics[scale=0.35,trim={0.5cm 0.5cm 0.5cm 0.5cm},clip]{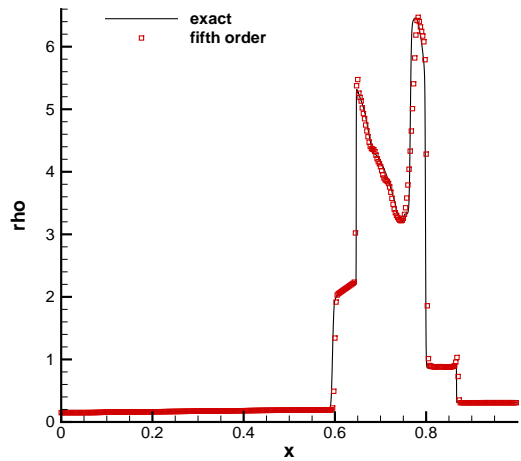}}
    \subfigure[$\mathrm{P_{6}T_{3}}$]{\includegraphics[scale=0.35,trim={0.5cm 0.5cm 0.5cm 0.5cm},clip]{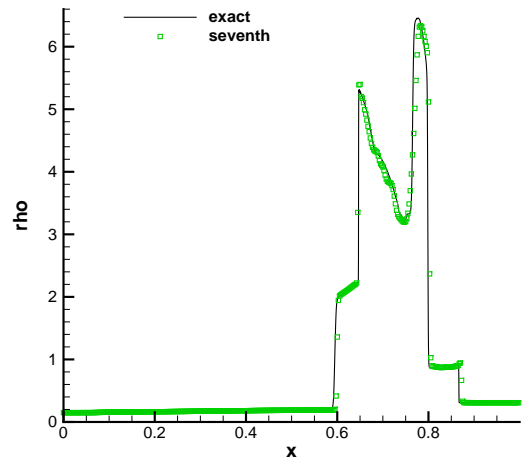}}
    \subfigure[$\mathrm{P_{8}T_{3}}$]{\includegraphics[scale=0.35,trim={0.5cm 0.5cm 0.5cm 0.5cm},clip]{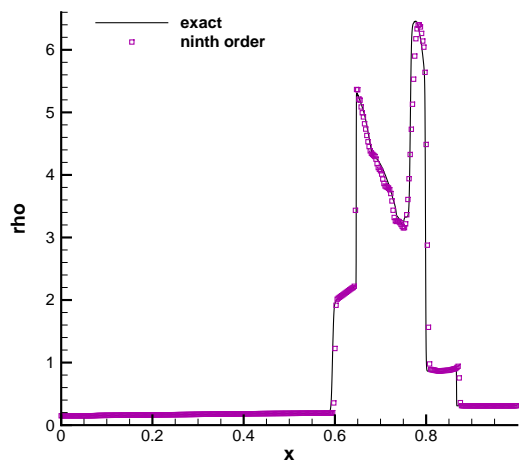}}
    \subfigure[$\mathrm{P_{10}T_{3}}$]{\includegraphics[scale=0.35,trim={0.5cm 0.5cm 0.5cm 0.5cm},clip]{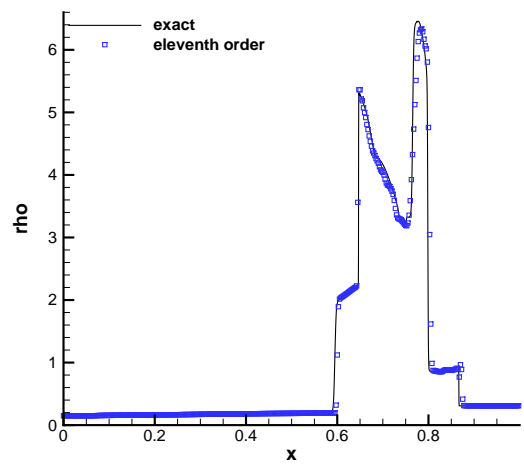}}
    \end{center}
	\protect\caption{Numerical results of two interacting blast waves problem for density field at t = 0.038 with 400 cells. \label{fig:blast}}

\end{figure}

Reflective boundary conditions are imposed at the two ends of computational domain. 
Two blast waves are generated by the initial jumps and then evolve, associated with violent  interactions among different flow structures. 
We used $400$ mesh cells as used in most literature for this test problem, and  depict the numerical density at time $t = 0.038$ in Fig.~\ref{fig:blast} against a reference solution 
obtained with WENO scheme using a very fine mesh.
The solution produced by high order WENO type schemes on $400$-cell mesh can be found in many published works, such as  \cite{jiang96,wenoz} where contact discontinuities are smeared significantly, especially the left-most contact discontinuity around $x\simeq 0.6$. On the contrary, the proposed $\mathrm{P}_{n}\mathrm{T}_{m}-\mathrm{BVD}$ schemes show overall better resolution and can capture the left-most contact discontinuity with only three points.  

\subsection{Accuracy test for 2D Euler equations}
We use a simple test case of two dimensional nonlinear Euler equations to show that the $\mathrm{P}_{n}\mathrm{T}_{m}-\mathrm{BVD}$ schemes will select the high-order interpolations of polynomials of linear weights for smooth region and realize high-order convergence rates in multi-dimensional implementations. It should be noted here that in order to realize truly high-order schemes for multi-dimensional Euler equations, sophisticated techniques introduced in \cite{TitarevToro,2dH1,2dH2} should be adopted. Nevertheless, for the sake of simplicity, we follow  \cite{veryhigh,yan} and use simple dimension-wise implementation for 2D Euler equations.  As reported in \cite{2dH2}, this simple multi-dimensional implementation is able to give adequate accuracy for problems involving shock waves. To achieve truly high order in multi-dimensions, $\mathrm{P}_{n}\mathrm{T}_{m}-\mathrm{BVD}$ schemes can be implemented in fashion of any high-order finite volume scheme as addressed in \cite{TitarevToro,2dH1,2dH2}.

In this test, the  initial velocity and pressure were specified uniform throughout the whole computational domain, while a sinusoidal perturbation was given to the density field as 
\begin{equation}
\rho_{0}(x,y)=1+0.5\sin(2\pi(x+y)),~~
u_{0}(x,y)=1.0,~~
v_{0}(x,y)=1.0,~~
p_{0}(x,y)=1.0.
\end{equation}
The computational domain was $[-1,1]\times[-1,1]$. We used gradually refined grids with periodic boundaries and obtained the numerical results at $t=2.0$. The numerical errors and convergence rates obtained by the $\mathrm{P}_{n}\mathrm{T}_{m}-\mathrm{BVD}$ schemes and the corresponding high order upwind schemes were summarized in Table~\ref{Tab:rate2DEuler} and Table~\ref{Tab:rate2DEulerUW} respectively. It is found that the $\mathrm{P}_{n}\mathrm{T}_{m}-\mathrm{BVD}$ schemes produce the same numerical errors and convergence rates as their corresponding upwind schemes. Thus, it is verified that  the $\mathrm{P}_{n}\mathrm{T}_{m}-\mathrm{BVD}$ schemes retrieve the linear high-order interpolation and achieve expected order accuracy in smooth regions for 2D Euler equations. 
    
\begin{table}[]
	\centering
	\caption{Numerical errors and convergence rate for 2D Euler equation. Results are given by  $\mathrm{P}_{n}\mathrm{T}_{m}-\mathrm{BVD}$ schemes.}
	\label{Tab:rate2DEuler}
	\begin{tabular}{l|lllll}
		\hline
		Schemes                                      & Mesh & $L_{1}$ errors & $L_{1}$ order & $L_{\infty}$ errors & $L_{\infty}$ order \\ \hline
		\multirow{4}{*}{$\mathrm{P_{4}T_{2}}$}
& $10\times10$ & $1.292\times10^{-1}$ &  & $1.996\times10^{-1}$ & \\		                     
& $20\times20$   & $1.076\times10^{-2}$ & 3.58 & $1.663\times10^{-2}$ & 3.58\\
& $40\times40$   & $3.917\times10^{-4}$ & 4.78 & $6.056\times10^{-4}$ & 4.78 \\
& $80\times80$  & $1.266\times10^{-5}$ & 4.95 & $1.985\times10^{-5}$ & 4.93 \\ \hline		
		\multirow{4}{*}{$\mathrm{P_{6}T_{3}}$}
& $10\times10$ & $5.094\times10^{-2}$ &  & $7.872\times10^{-2}$ & \\		                     
& $20\times20$   & $8.953\times10^{-4}$ & 5.83 & $1.383\times10^{-3}$ & 5.83\\
& $40\times40$   & $8.190\times10^{-6}$ & 6.77 & $1.269\times10^{-5}$ & 6.76 \\
& $80\times80$  & $6.653\times10^{-8}$ & 6.94 & $1.041\times10^{-7}$ & 6.92 \\ \hline
		\multirow{4}{*}{$\mathrm{P_{8}T_{3}}$}
& $10\times10$ & $1.701\times10^{-2}$ &  & $2.629\times10^{-2}$ & \\		                     
& $20\times20$   & $7.626\times10^{-5}$ & 7.80 & $1.178\times10^{-4}$ & 7.80\\
& $40\times40$   & $1.782\times10^{-7}$ & 8.74 & $2.760\times10^{-7}$ & 8.73 \\
& $80\times80$  & $3.641\times10^{-10}$ & 8.93 & $5.700\times10^{-10}$ & 8.91 \\ \hline
		\multirow{4}{*}{$\mathrm{P_{10}T_{3}}$}
		& $10\times10$ & $5.490\times10^{-3}$ &  & $8.483\times10^{-3}$ & \\		                     
& $20\times20$   & $6.631\times10^{-6}$ & 9.69 & $1.024\times10^{-5}$ & 9.69\\
& $40\times40$   & $3.967\times10^{-9}$ & 10.70 & $6.141\times10^{-9}$ & 10.70 \\
& $80\times80$  & $2.031\times10^{-12}$ & 10.93 & $3.188\times10^{-12}$ & 10.91 \\ \hline
	\end{tabular}
\end{table}

\begin{table}[]
	\centering
	\caption{Same as table \ref{Tab:rate2DEuler}, but by the high order linear upwind schemes.}
	\label{Tab:rate2DEulerUW}
	\begin{tabular}{l|lllll}
	\hline
	Schemes                                      & Mesh & $L_{1}$ errors & $L_{1}$ order & $L_{\infty}$ errors & $L_{\infty}$ order \\ \hline
	\multirow{4}{*}{5th order UW}
	& $10\times10$ & $1.292\times10^{-1}$ &  & $1.996\times10^{-1}$ & \\		                     
	& $20\times20$   & $1.076\times10^{-2}$ & 3.58 & $1.663\times10^{-2}$ & 3.58\\
	& $40\times40$   & $3.917\times10^{-4}$ & 4.78 & $6.056\times10^{-4}$ & 4.78 \\
	& $80\times80$  & $1.266\times10^{-5}$ & 4.95 & $1.985\times10^{-5}$ & 4.93 \\ \hline		
	\multirow{4}{*}{7th order UW}
	& $10\times10$ & $5.094\times10^{-2}$ &  & $7.872\times10^{-2}$ & \\		                     
	& $20\times20$   & $8.953\times10^{-4}$ & 5.83 & $1.383\times10^{-3}$ & 5.83\\
	& $40\times40$   & $8.190\times10^{-6}$ & 6.77 & $1.269\times10^{-5}$ & 6.76 \\
	& $80\times80$  & $6.653\times10^{-8}$ & 6.94 & $1.041\times10^{-7}$ & 6.92 \\ \hline
	\multirow{4}{*}{9th order UW}
	& $10\times10$ & $1.701\times10^{-2}$ &  & $2.629\times10^{-2}$ & \\		                     
	& $20\times20$   & $7.626\times10^{-5}$ & 7.80 & $1.178\times10^{-4}$ & 7.80\\
	& $40\times40$   & $1.782\times10^{-7}$ & 8.74 & $2.760\times10^{-7}$ & 8.73 \\
	& $80\times80$  & $3.641\times10^{-10}$ & 8.93 & $5.700\times10^{-10}$ & 8.91 \\ \hline
	\multirow{4}{*}{11th order UW}
	& $10\times10$ & $5.490\times10^{-3}$ &  & $8.483\times10^{-3}$ & \\		                     
	& $20\times20$   & $6.631\times10^{-6}$ & 9.69 & $1.024\times10^{-5}$ & 9.69\\
	& $40\times40$   & $3.967\times10^{-9}$ & 10.70 & $6.141\times10^{-9}$ & 10.70 \\
	& $80\times80$  & $2.031\times10^{-12}$ & 10.93 & $3.188\times10^{-12}$ & 10.91 \\ \hline
\end{tabular}
\end{table}

\section{Concluding remarks \label{sec:conclusion}}
In this study, a new type of very high order reconstruction schemes, so-called $\mathrm{P}_{n}\mathrm{T}_{m}-\mathrm{BVD}$, is proposed in the finite volume framework. The new scheme hybridizes linear-weight polynomial of degree $n$ and THINC function with $m$-level steepness. The effective reconstruction function in each cell is determined by the BVD algorithm which minimizes the numerical dissipation. High order accuracy can be realized in an efficient and straightforward way by directly increasing the degree of polynomial. We extend the new scheme to eleventh order in present study. The spectral analysis and numerical tests show that the $\mathrm{P}_{n}\mathrm{T}_{m}-\mathrm{BVD}$ schemes can retrieve the underlying low-dissipation linear schemes over all wave numbers for smooth solutions. The benchmark tests verify that the proposed schemes can capture sharp discontinuities with numerical oscillations effectively suppressed. Moreover, as the order is increased, small-scale flow features can be resolved with much less numerical dissipation. Thus this work provides an innovative and practical alternative approach for spatial reconstructions of very high order for finite volume method to solve hyperbolic conservative systems that contain discontinuous and smooth solutions of various scales.

\section*{Acknowledgment}
This work was supported in part by the fund from JSPS (Japan Society for the Promotion of Science) 
under Grant Nos. 15H03916, 17K18838 and 18H01366. 



\end{document}